\documentclass[prb,aps,twocolumn]{revtex4}
\usepackage{graphicx}
\usepackage{verbatim}   
\usepackage{color}      
\usepackage{subfigure}  
\usepackage{amsmath} 
\usepackage{amssymb}
\newcommand{\pderiv}[2]{\frac{\partial #1}{\partial #2}}
\begin{document}

\title{Effect of temperature on quantum criticality in 
the frustrated two-leg Heisenberg ladder}
\author{Brandon W. Ramakko}
\author{Mohamed Azzouz}
\email[Electronic Address: ]{mazzouz@laurentian.ca}  
\affiliation{Laurentian University, Department of Physics,
Ramsey Lake Road, Sudbury, Ontario P3E 2C6, Canada.}

\date{February 16, 2007}

\begin{abstract}
The antiferromagnetic Heisenberg model on the two-leg ladder with exchange interactions along the chains, rungs, and diagonals is studied using the Jordan-Wigner transformation and bond-mean-field theory. The inclusion of all three couplings introduces frustration to the system and depending on their relative strengths the ladder can adopt one of three possible magnetically-disordered gapped states. The phase diagram found in this mean-field approach is in very good agreement with the one calculated by Weihong and colleagues using the Lanczos exact diagonalization method. By analyzing the ground-state energy we study quantum criticality when the coupling parameters are varied at zero temperature. We study the effect of temperature on the phase boundaries, and find that the system
shows thermally-induced criticality for some values of the rung and diagonal coupling constants. All the phase transitions encountered in this system occur between disordered phases, and are all caused by frustration.
\end{abstract}
\maketitle

\newpage

\section{Introduction}

Quantum criticality is a current highly debated issue in strongly correlated electron systems.\cite{sachdev1999} There are three kinds of strongly correlated systems. The first ones are those with only localized electrons like the quantum Heisenberg-type spin systems, where only the spin degrees of freedom contribute to the physical properties. The second ones are those systems where electrons are mobile, and both the spin and electron degrees of freedom are relevant like in the high-$T_C$ materials away from half-filling. The third type of systems are those fermionic systems with both localized and itinerant electrons like in the Kondo-type (heavy-fermion) systems. In this work, we develop an analytical approach to study the quantum criticality phenomenon in the frustrated antiferromagnetic (AF) two-leg Heisenberg ladder, and the effect of temperature on this criticality. Because, absolute zero temperature cannot be reached in any experiment, it is important to investigate such temperature dependence. In the absence of frustration, the two-leg ladder has been analyzed extensively both numerically and analytically.\cite{Dagotto} The case with second-nearest-neighbor interactions along the diagonals is of interest because this interaction adds frustration to the system, and there is a possibility that in real two-leg ladder materials it might be significant. Some examples of real two-leg ladder materials are SrCu$_2$O$_3$,\cite{Azuma} Cu$_2$(C$_5$H$_{12}$N$_{2}$)$_2$Cl$_4$,\cite{Hayward,Chaboussant} and La$_6$Ca$_8$Cu$_{24}$O$_{41}$.\cite{Imai,Jd11} Existing numerical data on the frustrated ladder indicate that when the diagonal interaction is varied the system undergoes a quantum phase transition. The main motivation for the present work is reaching better understanding of quantum criticality in the frustrated two-leg ladder from a microscopic point of view and study the effect of temperature on it. the method we develop is an analytical approach based on the Jordan-Wigner (JW) transformation. It is first tested at zero temperature by making sure it reproduces all the numerically derived exact existing results, then it is applied at finite temperature. 

\begin{figure}
		\includegraphics[height=2.1cm]{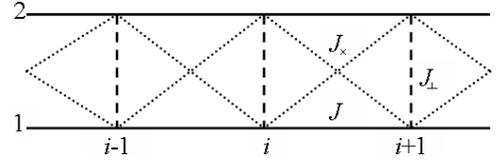}
		\centering
		\caption{The two-leg ladder showing the couplings along the chains, rungs, and diagonals is displayed.}
	\label{fig:Two leg ladder}
\end{figure}

The Hamiltonian for the spin-$\frac{1}{2}$ two-leg ladder with diagonal interactions is written as
\begin{eqnarray}
\label{eq:Ham}
H &=& J \sum_{i}^N \sum_{j=1}^2 \textbf{S}_{i,j}\cdot \textbf{S}_{i+1,j} +
J_\perp \sum_{i}^N \textbf{S}_{i,1}\cdot\textbf{S}_{i,2}\nonumber \\
&&+ J_{\times} \sum_{i}^N ( \textbf{S}_{i,1}\cdot\textbf{S}_{i+1,2} + \textbf{S}_{i+1,1}\cdot\textbf{S}_{i,2}),
\end{eqnarray}
where $J$ is the coupling along the chains, $J_\perp$ the transverse coupling, and $J_{\times}$ the coupling along the diagonals as seen in Fig.~\ref{fig:Two leg ladder}. The index $i$ labels the position of the spins along the two chains, each of which has $N$ sites. The first term sums the interactions of nearest-neighboring spins along the chains (legs) of the ladder, the second term sums the interactions of  the spins along the rungs, and the third term sums the interactions along the diagonals. As usual, $\textbf{S}_{i,j}$ is the spin operator.

The frustrated two-leg ladder has been studied numerically using Ising and dimer expansions,\cite{Jd6} Lanczos diagonalization technique,\cite{Jd11,Jd6,Jd4,Jd8} and density-matrix renormalization group (DMRG).\cite{Jd8,Jd3,Jd5,Jd10,white1996} It has also been studied analytically and we will next summarize briefly the different analytical theories that have been 
applied at zero temperature.
%
%
%
%
The (dimerized) valence-bond spin-wave theory by Xian\cite{Jd7}
was used to study the system. This theory breaks 
down for certain values of the couplings. For example it works
for $J/J_{\perp}<1/2$ (so in the strong coupling limit with $J_\perp\gg J$) when $J_{\times}/J_{\perp}=0$.
%
The abelian bosonization technique applied by Weighong {\it et al}.\cite{Jd6}, and the non-Abelian bosonization method applied by Allen and coworkers\cite{Jd13} work only in the the weak coupling 
limit $J_{\times},J_{\perp}<J$. The Lieb-Mattis theorem applied by Hakobyan\cite{hakobyan2007} puts limits on the transition line such as $J_{\perp}\leq 2J_{\times}$ but cannot determine its exact position. The reformulated weak-coupling field theory of Starykh and Balents\cite{starykh2004}
works in the limit $J_{\times},J_{\perp}<J$ only. These authors found that in the quantum model the classical transition 
at $J_{\perp}=2J_{\times}$ splits into two, implying the occurrance of a new 
phase, which was later disproved numerically.\cite{Jd8} The non-linear Sigma model was considered by Nedelcu {\it et al}.\cite{Jd12} who focused on the case where the diagonal couplings are different. However, for the case with equal diagonal interactions there are portions of the phase diagram where their theory fails. The non-perturbative effective field theory of Cabra {\it et al}.\cite{Jd9}
examined which spin bonds are the strongest in order to put limits on weak 
and strong coupling regimes. However, no zero-temperature phase diagram was produced in their work.

None of these analytical methods was able to create a complete phase diagram including the weak, intermediate, and strong coupling regimes which compares well with numerical data because of their various limitations. The exact numerical methods provide reliable information about the states of the system, but in order to understand even these numerical results one needs to develop analytical approaches, which are readily generalizable to finite temperature, and which apply to all coupling regimes. Using the bond-mean-field theory\cite{Azz1,Azz2,Azz3,Azz5} (BMFT) we seek better understanding of the phases of the frustrated two-leg ladder at both zero and nonzero temperatures. The BMFT is a mean-field theory that is based on the spin bond parameters. These parameters are not related in any way to 
any kind of long-range order. They are rather related to the spin-spin correlation function $\langle S_{i}^-S_{j}^+\rangle$, with $i$ and $j$ labelling adjacent sites in the direction where this correlation function is calculated. All quantities $\langle S_i^\alpha\rangle$, with $\alpha=x,\ y,\ z$, are zero in BMFT, implying the absence of any sort of long-range magnetic order.

This paper is organized as follows. In Sec.~\ref{sec:Method}, we explain how the BMFT, which is based on the two-dimensional (2D) JW transformation, is applied to the Hamiltonian for the two-leg ladder with interactions along the chains, rungs, and diagonals. The Hamiltonian is handled and decoupled similarly to that of Ref.~\cite{Azz3}. Quantum criticality, energy spectra, mean-field parameters, free energy, entropy, and specific heat are calculated in Sec. \ref{sec:Results}. Also, the zero and non-zero temperature phase diagrams are calculated. Criticality at non-zero temperature is examined in Sec.~\ref{sec:Tcriticality}. Discussion of our results is made in Sec.~\ref{sec:Discussion}. In Sec.~\ref{sec:Conclusion}, conclusions are drawn.

\section{Method}

\begin{figure}
		\includegraphics[height=4.50cm]{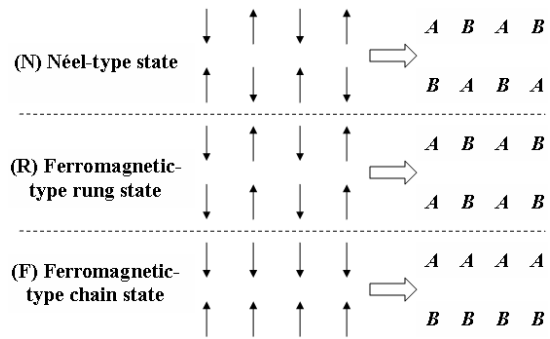}
		\centering
		\caption{In the left panel, the three possible ground states of the system in the Ising limit, 
		namely the N\'eel state, the ferromagnetic chain state, and the ferromagnetic rung state are drawn. In the right
		panel, the labeling of sublattices corresponding to the short-range spin orders that replace the long-range ones are
		shown for the Heisenberg limit.}
	\label{fig:Jd1}
\end{figure}

In the Ising limit, where quantum spin fluctuations are absent, the magnetically-ordered ground states are shown in the left panel of Fig. \ref{fig:Jd1}. Depending on the relative strength of the couplings, the system can be found in one of the three different ordered states displayed on the left of this figure. When $J_{\times}\ll J_{\perp}$, the system adopts the N\'eel state with ferromagnetic spin arrangements along the diagonals. When $J_{\times}\gg J_{\perp}$ the system adopts the ferromagnetic rung state. In this case, the AF arrangement shifts to the diagonals, and spins on the rungs are forced to adopt a ferromagnetic arrangement. When $J_{\times}\gg J$ and $J_{\perp}\gg J$, the system adopts a ferromagnetic chain state, where the spins on the chains are ordered ferromagnetically, and antiferromagnetically on the rungs.
The phase diagram in this classical limit will also be given later on for comparison.

In the Heisenberg limit with full quantum fluctuations, we assume that in a one-to-one correspondence each of the magnetic orders in Fig. \ref{fig:Jd1} evolves into a state which is characterized by only short-range order of the same kind as in the ordered state. So, we replace the rigid up and down orientations of the spins by sublattice labels $A$ and $B$, respectively, as shown in the right panel of Fig. \ref{fig:Jd1}. According to Ref.\cite{white1996}, the boundaries in the phase diagram of the Ising limit move as a consequence of quantum fluctuations, but the number of phases stays the same, namely three. For this reason, we continue to label the phases using the same terms as in the Ising limit except that we now add the word type to indicate that the phases are not magnetically ordered but are rather characterized by short-range order only. So, the phases are now called N\'eel-type (N-type) state, ferromagnetic-type rung (R-type) state, and ferromagnetic-type chain (F-type) state. For the N-type state in the absence of frustration, the ground-state of the two-leg ladder is a gapped disordered spin liquid dominated by short-range AF correlations for any nonzero $J_{\perp}$.\cite{Dagotto} The gap increases linearly with rung coupling, except near zero coupling where a slight downward curvature is present. The ferromagnetic-type rung state is also referred to as the Haldane-phase or rung triplet phase because in this case the system presents some resemblance to the $S=1$ chain. Precisely, in the case $J_\perp=0$ and $J_{\times}=J$, the Hamiltonian becomes $H=J\sum_i({\bf S}_{i,1}+{\bf S}_{i,2})\cdot({\bf S}_{i+1,1}+{\bf S}_{i+1,2})$, with low-lying excitations identical to those of the $S=1$ Haldane chain due to the fact that the interaction is now between the resultant of the two spins on the rungs.\cite{white1996} The N and F-type states are also known as the dimerized or singlet states because of the formation of singlets along the rungs in the quantum case. The two phases belong to the same universality class so they are only topologically distinct due to the symmetry of the system.\cite{Jd10,kim2000} For the N-type state, in the limit $J_\perp=\infty$ and $J_{\times}=0$, the ground state is made of independent rung singlets.

We use the spin arrangements on the right panel in Fig.~\ref{fig:Jd1} as a starting point. We assume that in the Heisenberg limit short-range AF correlations are important enough to justify its use. Note that these spin arrangements are not static in the quantum limit, but fluctuate while the relative average orientations of adjacent spins remain the same. We will confirm that the frustrated two-leg ladder is characterized by quantum criticality (criticality at zero temperature induced solely by quantum fluctuations), and show that it is also characterized by criticality at nonzero temperature. The finite-temperature phase transitions we propose for this practically one-dimensional system are not from order to disorder, but they are from a disordered phase to another disordered one; i.e., between states that differ by their short-range magnetic order only. 

\label{sec:Method}

\subsection{The JW transformation and BMFT}

The JW transformation for the two-leg Heisenberg ladder is defined as\cite{Azz3}
\begin{eqnarray}
\label{eq:2D JW}
S_{i,j}^{-} & = & c_{i,j}e^{i\phi_{i,j}}, \nonumber \\
S_{i,j}^{z} & = & n_{i,j}-1/2,\ \ \ \ 
n_{i,j}=  c_{i,j}^{\dag}c_{i,j}, \nonumber \\
\phi_{i,1} & = & \pi[\sum_{d=0}^{i-1}\sum_{f=1}^{2}n_{d,f}]\ \ {\rm\ for\ chain\ 1,} \nonumber \\
\phi_{i,2} & = & \pi[\sum_{d=0}^{i-1}\sum_{f=1}^{2}n_{d,f} + n_{i,1}] \ \ {\rm\ for\ chain\ 2}.
\end{eqnarray}
Here $i$ and $j$ are the coordinates along the chain and rung directions, respectively. The phases $\phi_{i,j}$ are chosen so that all the spin commutation relations are preserved. The $c_{i,j}^{\dag}$ operator creates a spinless fermion at site $(i,j)$, while $c_{i,j}$ annihilates one, and $n_{i,j}$ is the occupation number operator.

After applying the JW transformation~(\ref{eq:2D JW}) to the Hamiltonian~(\ref{eq:Ham}) we get 
\begin{eqnarray}
\label{eq:Ham1}
H & = &  H_{XY}
 + J \sum_{i}^N \sum_{j=1}^2 (c_{i,j}^{\dag}c_{i,j} - \frac{1}{2})(c_{i+1,j}^{\dag}c_{i+1,j} - \frac{1}{2}) \nonumber \\
&& + J_\perp \sum_{i}^N (c_{i,1}^{\dag}c_{i,1} - \frac{1}{2})(c_{i,2}^{\dag}c_{i,2} - \frac{1}{2}) \nonumber \\
&&+ J_{\times}\sum_{i}^{N} [(c_{i,1}^{\dag}c_{i,1} - \frac{1}{2})(c_{i+1,2}^{\dag}c_{i+1,2} - \frac{1}{2}) \nonumber \\
&& + (c_{i+1,1}^{\dag}c_{i+1,1} - \frac{1}{2})(c_{i,2}^{\dag}c_{i,2} - \frac{1}{2})],
\end{eqnarray}
where 
\begin{eqnarray}
\label{eq:XYHam} 
H_{XY} &=& 
\frac{J}{2} \sum_{i}^N [c_{i,1}^{\dag}e^{i\pi n_{i,2}}c_{i+1,1} + c_{i,2}^{\dag}e^{i\pi n_{i+1,1}}c_{i+1,2} + \textrm{H.c.}] \nonumber \\
&& + 
\frac{J_\perp}{2} \sum_{i}^N [c_{i,1}^{\dag}c_{i,2} + \textrm{H.c.}] \nonumber \\
&&+ \frac{J_{\times}}{2} \sum_{i}^N [c_{i,1}^{\dag}e^{i\pi (n_{i,2}+n_{i+1,1})}c_{i+1,2} + c_{i+1,1}^{\dag}c_{i,2} + \textrm{H.c.}]
\end{eqnarray}
is the $XY$ Hamiltonian of the frustrated two-leg ladder.
In BMFT, the interacting terms of the JW fermions are decoupled using the spin bond parameters. This approximation neglects fluctuations around the mean field points; $(O-\langle O\rangle)(O'-\langle O'\rangle) \approx 0$, where $O$ and $O'$ are any operators which are quadratic in $c^\dag$ and $c$.\cite{Azz2} This yields
\begin{equation}
\label{eq:HF}
OO' \approx \langle O\rangle O' + O\langle O'\rangle - \langle O\rangle \langle O'\rangle.
\end{equation}
To apply BMFT we introduce three mean-field bond parameters; $Q$ in the longitudinal direction, $P$ in the transverse direction, and $P'$ along the diagonal. These can be interpreted as effective hopping energies for the JW fermions~\cite{Azz1} in the longitudinal, transverse and diagonal directions, respectively:
\begin{eqnarray}
\label{eq:bondparameters}
Q & = & \langle c_{i,j}c_{i+1,j}^{\dag}\rangle, \nonumber \\
P & = & \langle c_{i,j}c_{i,j+1}^{\dag}\rangle,\nonumber \\
P' & = & \langle c_{i+1,j}c_{i,j+1}^{\dag}\rangle.
\end{eqnarray}
Keeping in mind that there is no long-range order~\cite{Mermin} so that $\langle S_{i,j}^{z}\rangle = \langle c_{i,1}^{\dag}c_{i,1}\rangle -1/2 = 0$, the Ising quartic terms in equation~(\ref{eq:Ham1}) can be decoupled and simplified using the Hartree-Fock approximation~(\ref{eq:HF}), and the bond parameters~(\ref{eq:bondparameters}) as follows:
\begin{eqnarray}
\label{eq:decouple Ising}
\left(c_{i,1}^{\dag}c_{i,1}\! - \!\frac{1}{2}\right)\left(c_{i+1,1}^{\dag}c_{i+1,1} \!-\! \frac{1}{2}\right) & \approx & 
 c_{i,1}^{\dag}c_{i+1,1}\langle c_{i,1}c_{i+1,1}^{\dag}\rangle 
 \nonumber \\ 
 &&+ \langle c_{i,1}^{\dag}c_{i+1,1}\rangle c_{i,1}c_{i+1,1}^{\dag} \nonumber \\
&& - \langle c_{i,1}^{\dag}c_{i+1,1}\rangle \langle c_{i,1}c_{i+1,1}^{\dag}\rangle \nonumber \\
& = & Qc_{i,1}^{\dag}c_{i+1,1} + Q^*c_{i+1,1}^{\dag}c_{i,1}\nonumber
\\ && + |Q|^2.
\end{eqnarray}
Note that when decoupled in the magnetization channel, the quartic terms give
%
%
$
\left(c_{i,1}^{\dag}c_{i,1} - \frac{1}{2}\right)\left(c_{i+1,1}^{\dag}c_{i+1,1} - \frac{1}{2}\right) 
 \approx 0,
$
%
%
which is a consequence of the absence of magnetic long-range order.
Then, the Hamiltonian becomes
\begin{eqnarray}
\label{eq:Ham2}
H & = & H_{XY} 
+ J \sum_{i}^N \sum_{j=1}^2 [ Qc_{i,j}^{\dag}c_{i+1,j} + Q^*c_{i+1,j}^{\dag}c_{i,j}] \nonumber \\
&& + J_\perp \sum_{i}^N [Pc_{i,1}^{\dag}c_{i,2} + P^*c_{i,2}^{\dag}c_{i,1}] \nonumber \\
&& +J_{\times}\sum_{i}^{N} [ P'c_{i,1}^{\dag}c_{i+1,2}+P'^*c_{i+1,2}^{\dag}c_{i,1}\nonumber \\
&&+ P'c_{i+1,1}^{\dag}c_{i,2}+P'^*c_{i,2}^{\dag}c_{i+1,1}] \nonumber \\
&&  +2NJ_{\times}|P'|^{2} + 2NJ|Q|^2 + NJ_{\perp}|P|^2.
\end{eqnarray}
Next, we write this Hamiltonian using the three different spin configurations in the right panel of Fig. \ref{fig:Jd1}. These configurations are instantaneous (not static) configurations in which adjacent spins in any direction keep on average the same relative orientations with respect to each other, but fluctuate globally on a time scale determined by the strongest coupling constant so that any kind of long-range magnetic order is absent. These fluctuations are a consequence of the quantum fluctuations. The three competing configurations of Fig. \ref{fig:Jd1} lead to three different quantum gapped spin liquid states, each characterized by its own short-range spin correlations and symmetry.

\subsection{N\'eel-type State}
\label{sec:Neel}

\begin{figure}
		\includegraphics[height=2.1cm]{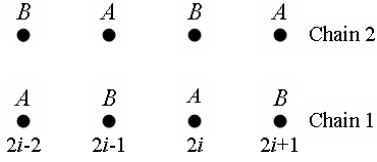}
		\centering
		\caption{The lattice is subdivided into two sublattices in the case of the N-type state.}
	\label{fig:lattice1}
\end{figure}

In the N-type state, the spin arrangement at any time consists on average of antiparallel spins on adjacent sites in both the longitudinal and rung directions; the spins on the diagonals thus prefer to align parallel to each other. In the limit $J_{\times} \ll J,J_{\perp}$, spins tend to form loose spin singlets on adjacent sites along both the chains and rungs if $J_{\perp}\sim J$ but strong spin singlets on the rungs if $J_{\perp} \gg J \gg J_{\times}$. The word N\'eel is not used to indicate long-range order but only the fact that the spins orientations show short AF order. For this reason, we divide the lattice into two sublattices; there are $c_{i,j}^{A}$ fermions on sublattice $A$, and $c_{i',j'}^{B}$ on sublattice $B$, where $(i',j')$ is any adjacent site to $(i,j)$, Fig.~\ref{fig:lattice1}. 
\begin{figure}
		\includegraphics[height=2.1cm]{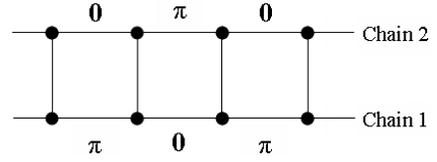}
		\centering
		\caption{The phase $\pi$ alternates along the chains and is zero everywhere else. The flux per plaquette is 
		$\pi$ on average.\cite{Affleck}}
	\label{fig:plaquette}
\end{figure}

Following Ref.\cite{Affleck}, we set the average phase per plaquette to be $\pi$. We choose the configuration seen in figure~\ref{fig:plaquette} which was suggested by Azzouz \textit{et al.} in Ref.~\cite{Azz3}. In the latter, this configuration is used to get rid of the phase terms in the $XY$ Hamiltonian $H_{XY}$; the only effect is that the sign of the hopping term in the JW XY term becomes alternated along the chains. For the Ising term, we set $Q_{i,j} = Qe^{i\phi_{i,j}}$ where $Q$ is site independent.\cite{Azz5} Here $\phi_{i,j}$ is the phase of the bond corresponding to Fig.~\ref{fig:plaquette} such that $\phi = \pi$ or $0$ along the chains, which means that the $Q$ terms alternate sign along the chains just like the $XY$ terms do. This is necessary in order to recover the proper result in the limit $J_{\times}$ and $J_{\perp}$ go to 0, in which we get a result comparable to that of des Cloiseaux and Pearson~\cite{Cloiseaux} for the spin excitation spectrum for a single Heisenberg chain, namely
%
%
$
E(k) = \frac{\pi}{2}J\left|\sin k\right|.
$
%
%
The alternating sign for $Q$ can be justified in the same way as in Ref.~\cite{Azz5}. 

A 1D Fourier transform along the chains is performed on the Hamiltonian keeping the chain index in the real space. 
The mean-field Hamiltonian can be expressed in the form
\begin{equation}
H = \sum_{k}\Psi_{k}^{\dag}\mathcal{H}\Psi_{k}+ 2NJQ^2 + NJ_{\perp}P^{2}+ 2NJ_{\times}P'^2
\label{mfhamiltonian}
\end{equation} 
with the Nambu spinor defined by
\begin{equation}
\Psi^{\dag} = (c_{1k}^{A\dag}\ c_{1k}^{B\dag}\ c_{2k}^{A\dag}\ c_{2k}^{B\dag}),
\end{equation}
and the Hamiltonian density given by
\begin{equation}
\label{densityNtype}
\mathcal{H} =
\begin{pmatrix}
0 & iJ_{1}\sin k & J_{\times1}\cos k & \frac{J_{\perp 1}}{2} \\
-iJ_{1}\sin k & 0 & \frac{J_{\perp 1}}{2} & J_{\times1}\cos k \\
J_{\times1}\cos k &  \frac{J_{\perp 1}}{2} & 0 & iJ_{1}\sin k \\
\frac{J_{\perp 1}}{2} & J_{\times1}\cos k & -iJ_{1}\sin k & 0
\end{pmatrix},
\end{equation}
with
\begin{eqnarray}
J_{1} &=&  J(1+2Q), \nonumber \\
J_{\perp1} &=&  J_{\perp}(1+2P), \nonumber \\
J_{\times1} &=&  J_{\times}(1+2P').
\end{eqnarray}
Diagonalizing $\mathcal{H}$ yields the energy eigenvalues $\pm E_{N1}$ and $\pm E_{N2}$ where
\begin{eqnarray}
E_{N1}(k) \!&\! =\! &\! J_{\times1}\cos k + \sqrt{J_{1}^{2}\sin^{2} k + \frac{J_{\perp 1}^{2}}{4}}, \nonumber \\
E_{N2}(k) \!&\! =\! & \!J_{\times1}\cos k - \sqrt{J_{1}^{2}\sin^{2} k + \frac{J_{\perp 1}^{2}}{4}}. 
\end{eqnarray}
Note that the subscript $N$ in $E_{Np}$ is used to remind ourselves of the N-type state.
Similarly, the eigenenergies of each of the remaining states will be labeled using its appropriate
subscript. The free energy corresponding to our Hamiltonian is
\begin{eqnarray}
\label{free energy}
F & = & JQ^2 + \frac{J_{\perp}P^{2}}{2} + J_{\times}P'^{2} \nonumber \\
&& - \frac{k_{B}T}{4N} \sum_{k}\sum_{s=\pm}\sum_{p=1,2}\ln[1 + e^{s\beta E_{Np}(k)}].
\end{eqnarray}
The parameters are determined by minimizing $F$ with respect to $Q$, $P$, and $P'$, a calculation which leads to the following set of self-consistent equations
\begin{eqnarray}
\label{eq:sc}
Q & = & \frac{1}{8NJ}\sum_{k}\sum_{p=1,2} \pderiv{E_{Np}(k)}{Q}\tanh\bigg[\frac{\beta E_{Np}(k)}{2}\bigg],  \nonumber \\
P & = & \frac{1}{4NJ_{\perp}}\sum_{k}\sum_{p=1,2} \pderiv{E_{Np}(k)}{P}\tanh\bigg[ \frac{\beta E_{Np}(k)}{2}\bigg],  \nonumber \\
P' & = & \frac{1}{8NJ_{\times}}\sum_{k}\sum_{p=1,2} \pderiv{E_{Np}(k)}{P'}\tanh\bigg[ \frac{\beta E_{Np}(k)}{2}\bigg]. 
\end{eqnarray}
The partial derivatives of the energies with respect to $Q$, $P$, and $P'$ are given by
\begin{eqnarray}
\label{partialdev}
\pderiv{E_{Np}}{Q} & = & \frac{(-1)^{p+1}2JJ_{1}\sin^{2}k}{\sqrt{J_{1}^{2}\sin^{2}k + \left(\frac{J_{\perp 1}}{2}\right)^{2}}}, \nonumber \\
\pderiv{E_{Np}}{P} & = &\! \frac{(-1)^{p+1}J_{\perp}J_{\perp 1}}{2\sqrt{J_{1}^{2}\sin^{2}k + \left(\frac{J_{\perp 1}}{2}\right)^{2}}}, \nonumber \\
\pderiv{E_{Np}}{P'} & = & 2J_{\times}\cos k, \qquad \qquad \textrm{with $p=1,2$}.
\end{eqnarray}
Next, we will analyze the F-type chain state.

\subsection{Ferromagnetic-type chain state}

\begin{figure}
		\includegraphics[height=4.0cm]{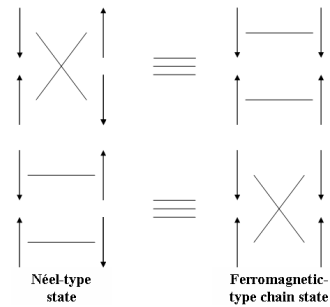}
		\centering
		\caption{The coupling along the diagonal (chain) in the N-type state is equivalent to the coupling along the chain (diagonal) in the F-type state.}
	\label{fig:Jd=J}
\end{figure}

In this state the instantaneous spin arrangement is AF along the diagonals and rungs, but ferromagnetic along the chains. The key thing to realize is the fact that the Hamiltonian is symmetric with respect to exchanging two spins along the rungs at even sites.\cite{Jd6,Jd3} So we will get similar spectra if the diagonal terms alternate sign and the terms along the chains have all the same sign (phase $\pi$ per plaquette). The spectra are the same as for the N-type state except that $J_1$ and $J_{\times1}$ are exchanged. Figure~\ref{fig:Jd=J} illustrates this correspondence.

A similar calculation was done by Xi Dai and Zhao-Bin Su~\cite{XiDai} for the two-leg ladder without diagonal interactions. They chose to alternate the index of the legs along the chain. They argued that this is justifiable because it still preserves the commutation relation when they use the JW transformation and because it gives them the expected results. If you twist their ladder to place the sites that are labeled 1 and 2 so that they each form a leg of the ladder you will realize that instead of calculating the interaction along the legs of the ladder in the N-type state they actually consider the interaction along the diagonal for the F-type state.

In their mean-field approach they end up with better results at $J_{\perp}/J=1$ (and $J_{\times}/J=0$) than with the approach of Azzouz {\it et al.}~\cite{Azz3} which we are using here. We are however not using their approach because it does not describe properly the limit $J_{\perp}/J \to 0$. On the contrary, the method we use here describes well this limit. The right description of this limit is crucial for any investigation of the phase diagram.

\begin{figure}
		\includegraphics[height=2.1cm]{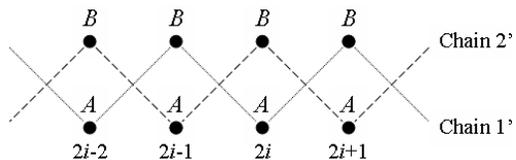}
		\centering
		\caption{The lattice is drawn in the case of the F-type state.}
	\label{fig:lattice2}
\end{figure}
The bipartite lattice for this state can be seen in Fig.~\ref{fig:lattice2}. 
We define two new chains 1' and 2' which are obtained by relabeling the sites along the diagonals as indicated in Fig.~\ref{fig:lattice2}. In this way the Hamiltonian density we get for the F-type state has the same expression as the one of the N-type state (\ref{densityNtype}) with $J_1$ replaced by $J_{\times1}$ and vice versa.
%
%
%
%
%
%
%
%
%
Explicitly, the energy eigenvalues are now $\pm E_{F1}$ and $\pm E_{F2}$, with
\begin{eqnarray}
E_{F1}(k) \!&\! =\! &\! J_{1}\cos k + \sqrt{J_{\times1}^{2}\sin^{2}k + \frac{J_{\perp 1}^{2}}{4}}, \nonumber \\
E_{F2}(k) \!&\! =\! & \!J_{1}\cos k - \sqrt{J_{\times1}^{2}\sin^{2}k + \frac{J_{\perp 1}^{2}}{4}}.
\end{eqnarray}
The equations for the free energy and the mean-field parameters have the same forms as for the N-type state, 
(\ref{free energy}) and (\ref{eq:sc}), respectively, except that now the partial derivatives of 
the energies with respect to $Q$, $P$, and $P'$ are
\begin{eqnarray}
\pderiv{E_{Fp}}{Q} & = &  2J\cos k,\nonumber \\
\pderiv{E_{Fp}}{P} & = &\! \frac{(-1)^{p+1}J_{\perp}J_{\perp 1}}{2\sqrt{J_{\times1}^{2}\sin^{2} k + \frac{J_{\perp 1}^{2}}{4}}},  \\
\pderiv{E_{Fp}}{P'} & = & \frac{(-1)^{p+1}2J_{\times}J_{\times1}\sin^{2} k}{\sqrt{J_{\times1}^{2}\sin^{2} k + \frac{J_{\perp 1}^2}{4}}}, \qquad \qquad \textrm{with $p=1,2$}.\nonumber
\end{eqnarray}

\subsection{Ferromagnetic-type Rung State}

\begin{figure}
		\includegraphics[height=2.1cm]{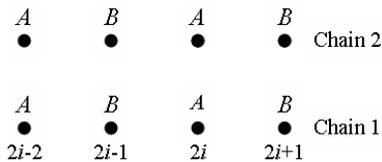}
		\centering
		\caption{The bipartite character of the lattice in the case of the R-type state is shown.}
	\label{fig:lattice3}
\end{figure}

This state is realized when the spins arrangement at any time is such that the spins along the diagonals are antiparallel and the spins along the rungs are parallel.
The bipartite lattice for this state is displayed in Fig.~\ref{fig:lattice3}. In this state both the diagonal and the chain terms link $A$ and $B$ fermions and vice versa; i.e., they are AF. Based on our previous arguments in Sec.~\ref{sec:Neel}, a phase of $\pi$ per plaquette can be used, with the mean-field phase this time alternating between $0$ and $\pi$ along the chains like in the N-type state.

A 1D Fourier transform is performed along the chains while keeping the chain labels in real space. The Mean-field Hamiltonian has the same expression as Eq.~(\ref{mfhamiltonian})
%
%
%
%
%
%
with the Nambu spinor now defined by
\begin{equation}
\Psi^{\dag} = ( c_{1k}^{A\dag} \ c_{2k}^{A\dag} \ c_{1k}^{B\dag} \ c_{2k}^{B\dag} )
\end{equation}
and the Hamiltonian density given by
\begin{equation}
\mathcal{H} =
\begin{pmatrix}
0 & \frac{J_{\perp 1}}{2} & iJ_{1}\sin k & J_{\times1}\cos k \\
\frac{J_{\perp 1}}{2} & 0 & J_{\times1}\cos k & iJ_{1}\sin k \\
-iJ_{1}\sin k & J_{\times1}\cos k & 0 & \frac{J_{\perp 1}}{2} \\
J_{\times1}\cos k & -iJ_{1}\sin k & \frac{J_{\perp 1}}{2} & 0
\end{pmatrix}.
\end{equation}
Diagonalizing this matrix yields the energy eigenvalues $\pm E_{R1}$ and $\pm E_{R2}$ with
\begin{eqnarray}
E_{R1}(k) \!&\! =\! &\!  \frac{J_{\perp 1}}{2}+ \sqrt{J_{1}^{2}\sin^{2}k + J_{\times1}^{2}\cos^{2}k}, \cr 
E_{R2}(k) \!&\! =\! & \! \frac{J_{\perp 1}}{2}- \sqrt{J_{1}^{2}\sin^{2}k + J_{\times1}^{2}\cos^{2}k}.
\end{eqnarray}
gain, the equations for the free energy and the mean-field parameters have the same form as for the N-type state, with the partial derivatives of the energies with respect to $Q$, $P$, and $P'$ replaced by
\begin{eqnarray}
\pderiv{E_{Rp}}{Q} & = &  \frac{(-1)^{p+1}2JJ_{1}\sin^{2}k}{\sqrt{J_{1}^{2}\sin^{2}k + J_{\times1}^{2}\cos^{2}k}},\nonumber \\
\pderiv{E_{Rp}}{P} & = &\! J_{\perp},  \\
\pderiv{E_{Rp}}{P'} & = & \frac{(-1)^{p+1}2J_{\times}J_{\times1}\cos^{2}k}{\sqrt{J_{1}^{2}\sin^{2}k + J_{\times1}^{2}\cos^{2}k}}, \qquad \qquad \textrm{with $p=1,2$}.\nonumber
\end{eqnarray}
Now that we have derived the mean-field equations for all three states, we solve them in order to get the zero-temperature and temperature-dependent phase diagrams. 
These equations are solved numerically, except in the high-temperature limit where they are solved both analytically and numerically. Our results will be compared with existing exact numerical data.

\section{Results}

\label{sec:Results}

\subsection{Zero-temperature phase diagram}
\label{sec:zerotemp}

\begin{figure}
\begin{center}
\subfigure{
\includegraphics[scale=0.22]{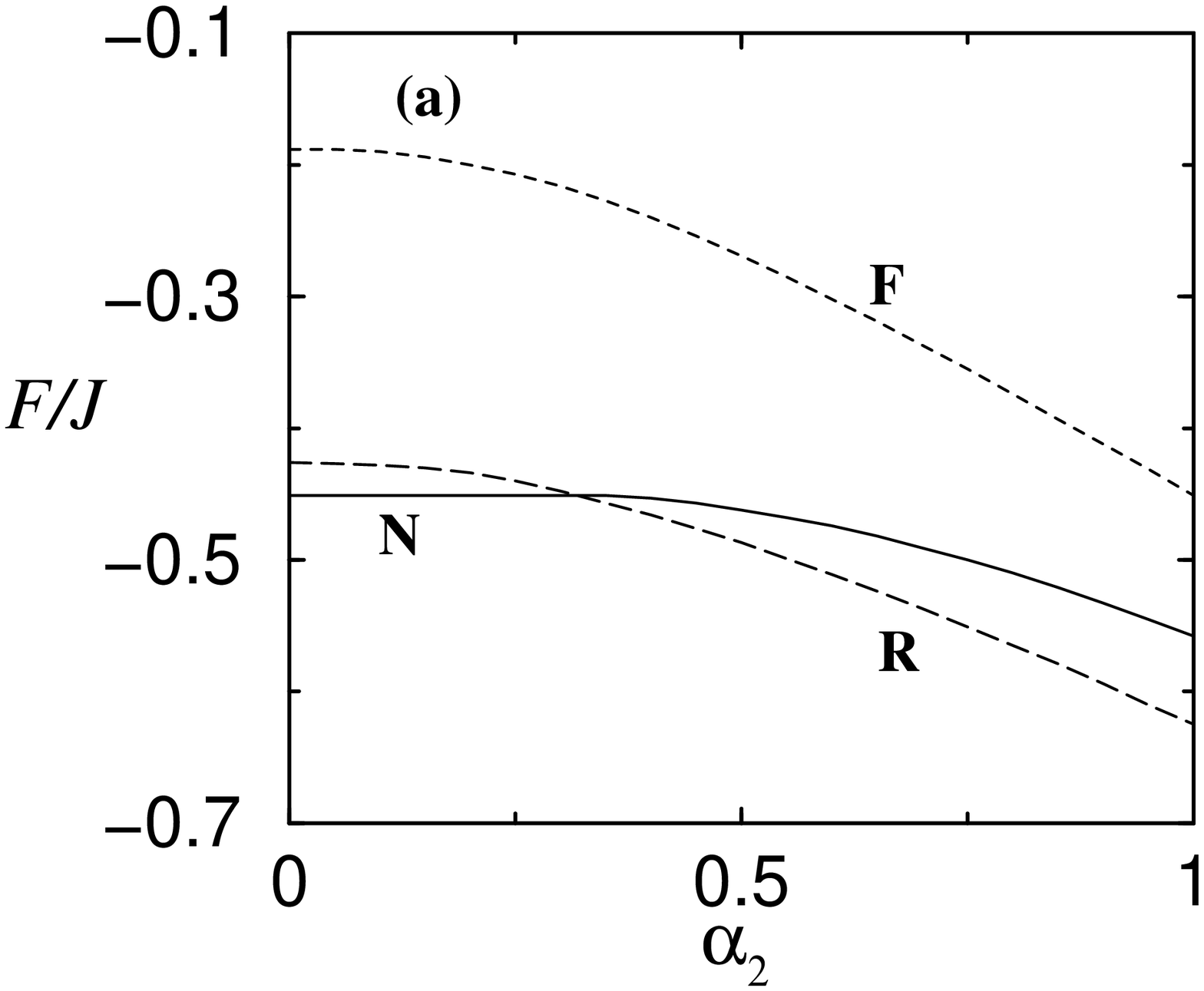}}
\subfigure{
\includegraphics[scale=0.22]{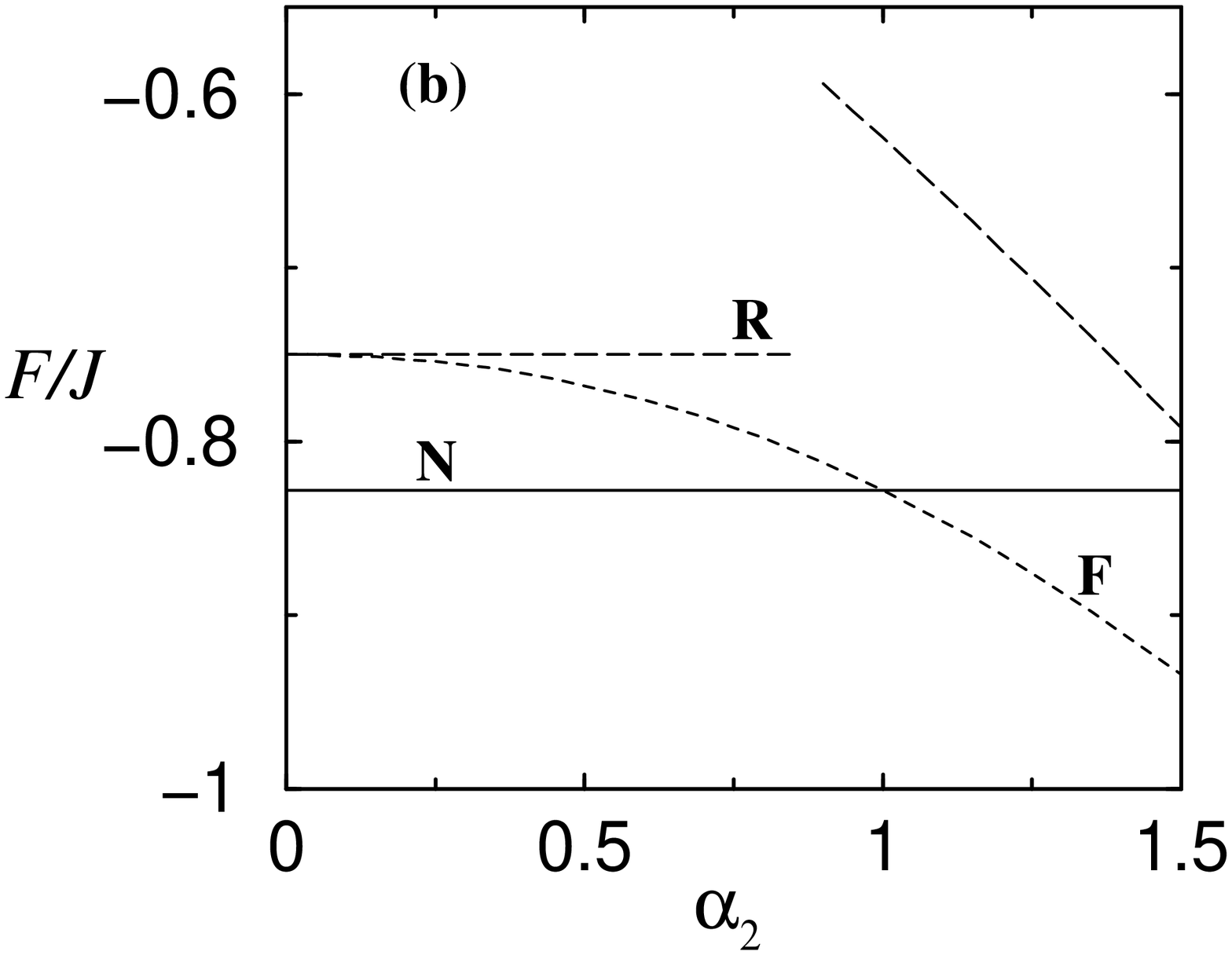}}
\caption{\label{fig:freeEnergies} The zero-$T$ free (ground-state) energies as calculated by BMFT for the N-type state (N), R-type state (R), and the F-type state (F) are plotted versus $\alpha_2 = J_{\times}/J$. (a) $\alpha_1 = J_{\perp}/J = 0.5$; there is a transition from the N-type state to the R-type state. (b) $\alpha_1 = 2$; there is a transition from the N-type state to the F-type state. There is a discontinuity in the free energy of the R-type state due to a sudden change in the bond parameters. This is of no interest because the only stable state is the one with the lowest free energy, and the transition is determined by the crossing of the lower free energies.}
\end{center}
\end{figure}

\begin{figure}
		\includegraphics[height=4.0cm]{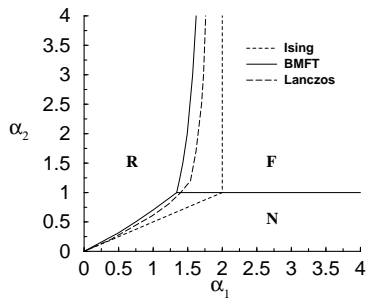}
		\centering
		\caption{The $(\alpha_1,\alpha_2)$-phase diagram we calculate is shown. It is compared to the Lanczos-method,\cite{Jd6} and Ising Expansion\cite{Jd6} results. The boundaries are between the N-type state (N), R-type state (R), and the F-type state (F). For comparison, the phase boundaries (dotted lines) in the Ising limit are also shown.}
	\label{fig:phase1}
\end{figure}

 The free (ground-state) energies of all three states are calculated as functions of coupling constants and compared. From thermodynamic considerations the state with the lowest free energy is the stable one, and whenever free energies cross a phase transition takes place.  Since only the ratios of the couplings are important we define $\alpha_1 =J_{\perp}/J$ and $\alpha_2 = J_{\times}/J$. The calculation was carried out for different sets of values of $\alpha_1$ and $\alpha_2$, with $J$ being the unit of energy. We found that at some values of these couplings free energies cross, which means that a phase transition occurs. You can refer to Fig.~\ref{fig:freeEnergies} for a couple of examples. In this way we have obtained the phase diagram at zero temperature. The phase transitions found here using BMFT are first-order ones for all values of $\alpha_2$. This agrees relatively well with most of the work done thus far by numerical\cite{Jd4,Jd6} and analytical methods.\cite{Jd7} For small $\alpha_2$, most numerical methods lacked the required accuracy to determine the order of the transition, but DMRG calculation by Wang\cite{Jd3} found that for $\alpha_2 < 0.287$ the transition is of second-order character, and for all larger values it is first order. Because these transitions take place at zero $T$ as a consequence of varying the diagonal interaction, they can be labeled as quantum phase transitions.\cite{sachdev1999} Experimentally for a real material, one can vary the pressure and hope that the diagonal (or any other) coupling changes enough so that the critical region is reached.
The $(\alpha_1,\alpha_2)$-phase diagram
we calculated is compared to the Lanczos-technique data of Ref.\cite{Jd6} in Fig.~\ref{fig:phase1}.
The agreement between the Lanczos method data and our results is very good, a fact that indicates that the present mean-field treatment is acceptable. The line at $\alpha_2=1$ is exact and its placement is a consequence of the Hamiltonian symmetry with respect to exchanging $J$ and $J_{\times}$, which BMFT fully satisfies.

\subsection{Mean-field parameters}

The mean-field bond parameters $Q$, $P$, and $P'$ are not order parameters in the conventional way because they are not 
related in any way to any sort of long-range order. In BMFT, they 
are a measure of the AF fluctuations in the system. Each of these parameters can be interpreted as indicating the presence of a strong spin bond in the spatial direction in which this parameter is found to be significant. The spin bond consists of a renormalized spin singlet formed on adjacent lattice sites. It is therefore important to know the coupling dependence of these parameters. We found that it is the combination of how these parameters and free energies depend on coupling constants that determines the phase boundaries between the three possible states; N-type, R-type, and F-type. 
The zero-$T$ mean-field parameters are plotted in Fig.~\ref{fig:parameters,T=0} as functions of $\alpha_2$ and $\alpha_1$. From the analysis of free energy we determine the initial and final states as well as the transition points. In each state the parameter that is zero corresponds to the direction with ferromagnetic arrangement. For example, in Fig.~\ref{fig:parameters,T=0}(a) for small $\alpha_2$, we are in the N-type state with $P'=0$ and other parameters ($Q$ and $P$) hardly changing as $\alpha_2$ increases. At $\alpha_2 =0.7$, there is a transition to the R-type state, which is accompanied by a sharp change where $P$ vanishes and $P'$ increases sharply.  
\begin{figure}
\begin{center}
\subfigure{
\includegraphics[scale=0.22]{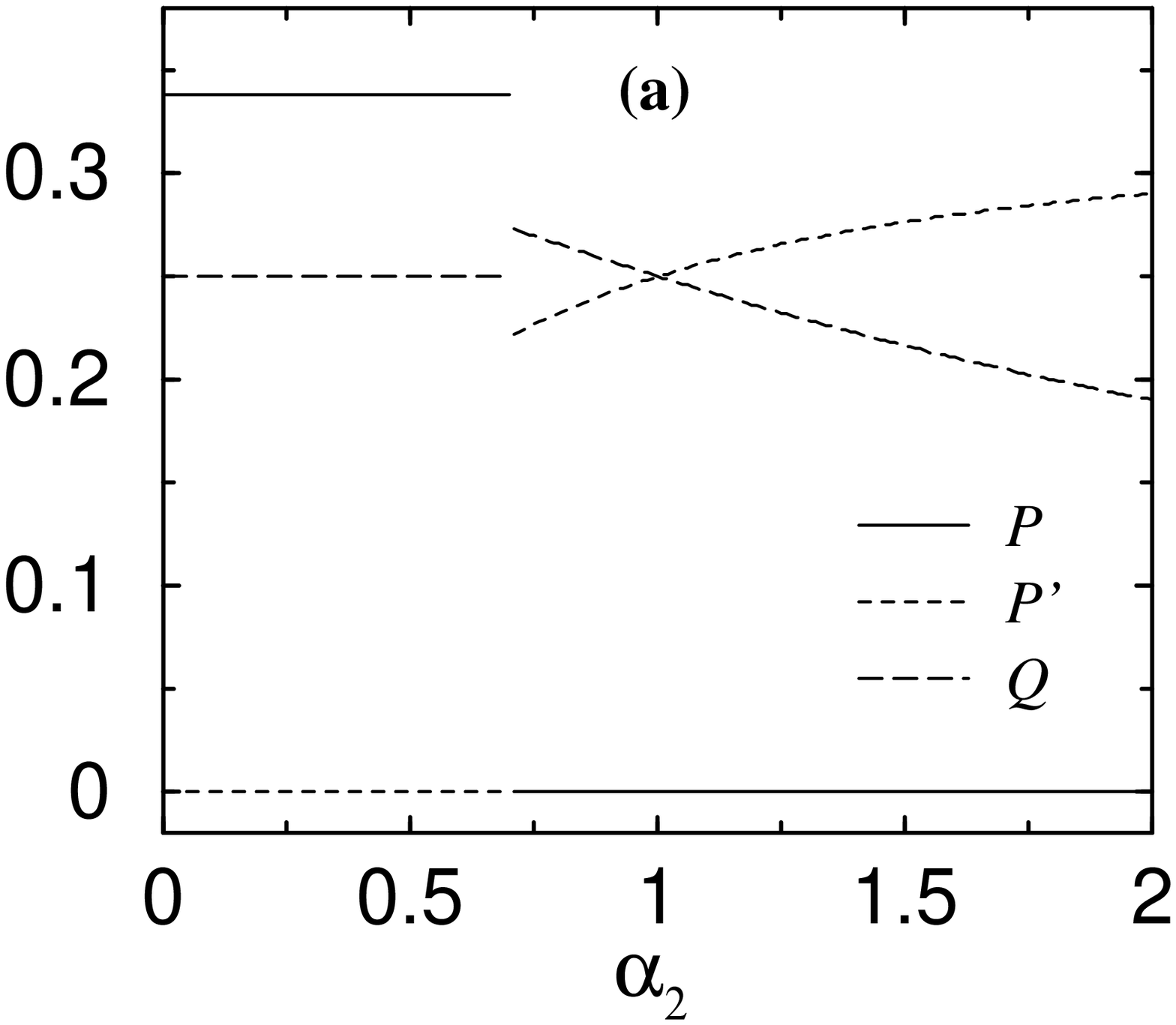}}
\subfigure{
\includegraphics[scale=0.22]{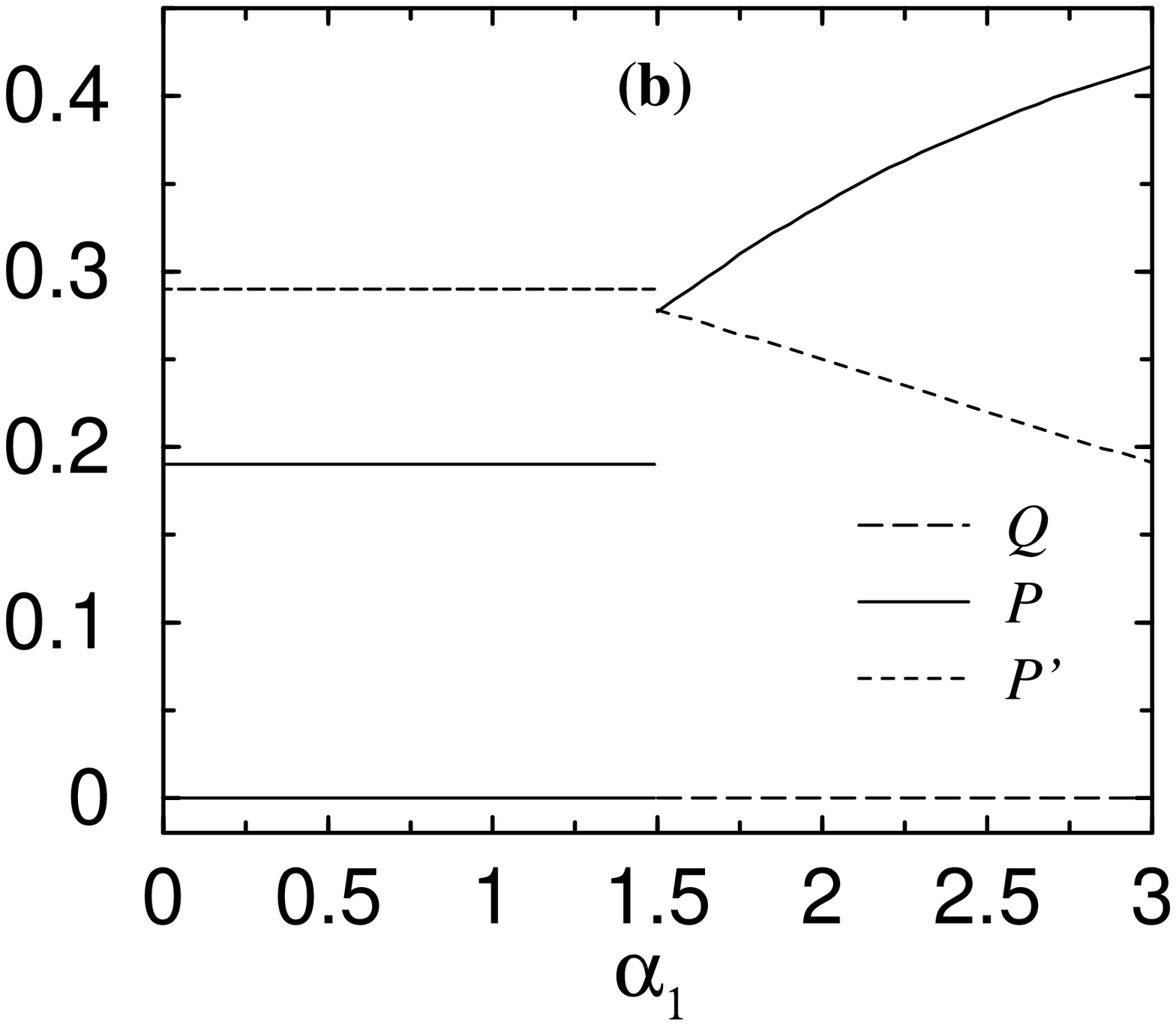}}
\caption{\label{fig:parameters,T=0} (a) The mean-field parameters are displayed as functions of $\alpha_2$ for $\alpha_1 =1$. There is a transition from the N-type state to the R-type state at $\alpha_2 =0.7$. (b) The mean-field parameters are displayed as functions of $\alpha_1$ for $\alpha_2 =2$. There is a transition from the R-type state to the F-type state at $\alpha_1 =1.5$.}
\end{center}
\end{figure}

\subsection{Energy gap and spectra}

\begin{figure}
\begin{center}
\subfigure{
\includegraphics[scale=0.22]{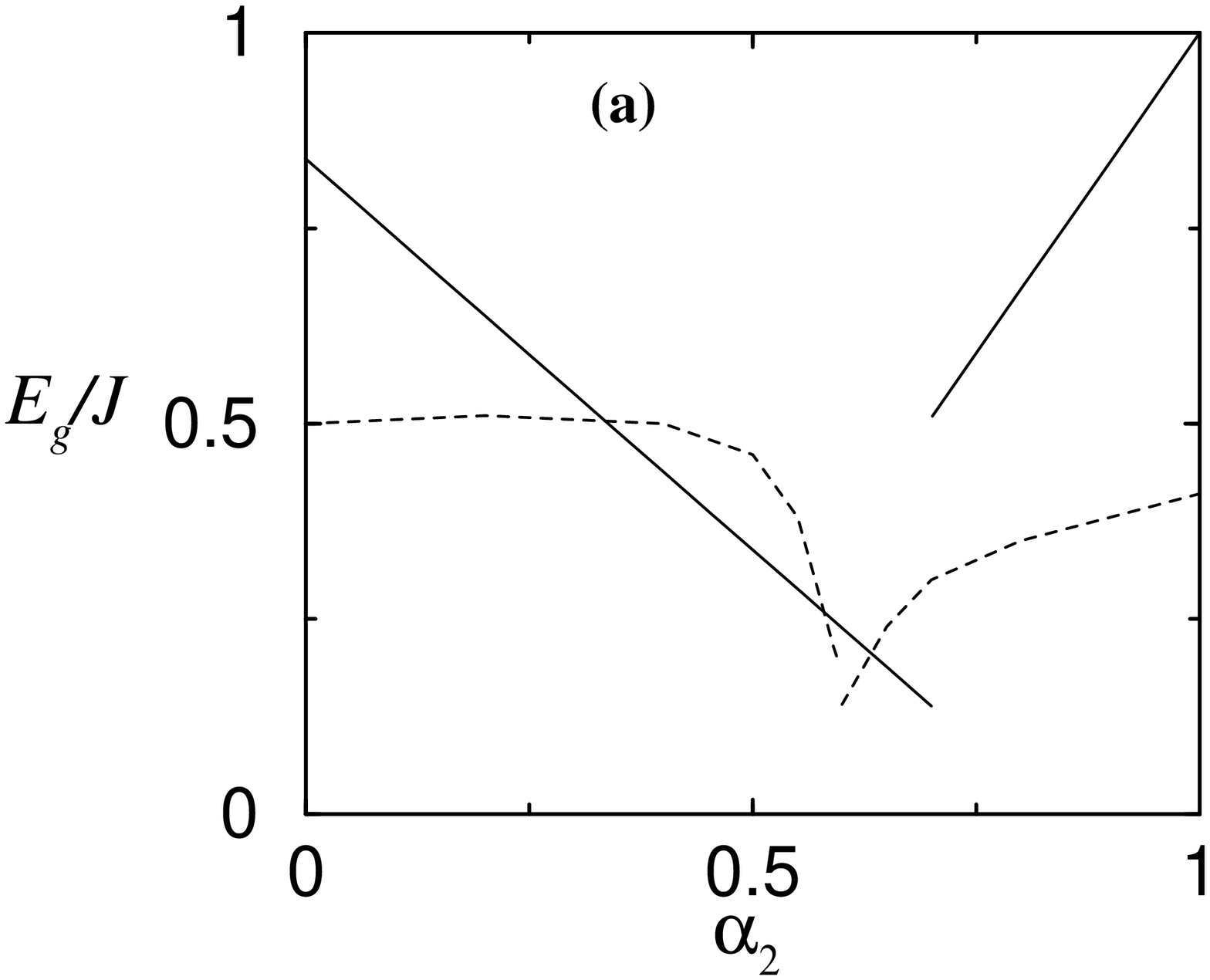}}
\subfigure{
\includegraphics[scale=0.22]{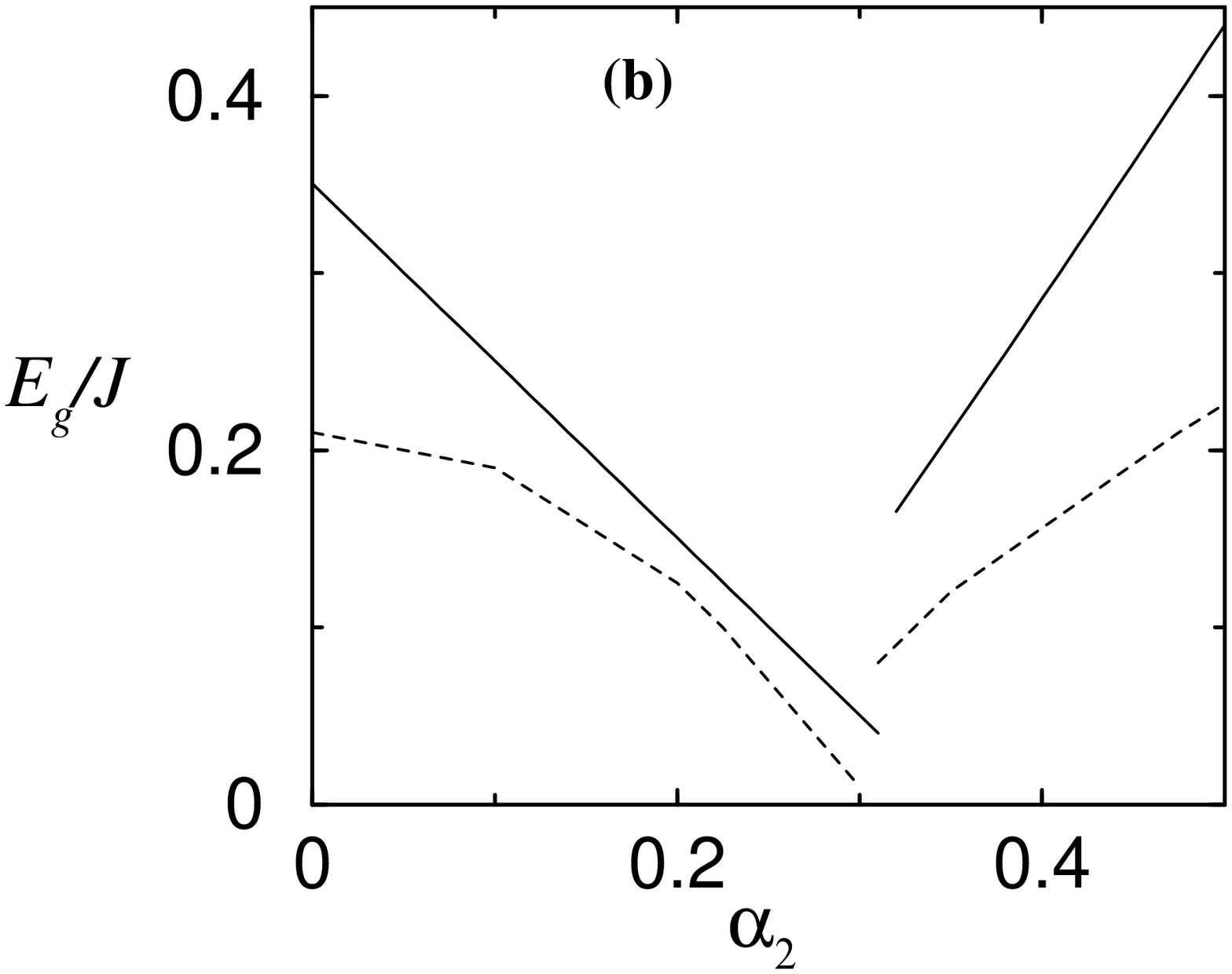}}
\subfigure{
\includegraphics[scale=0.22]{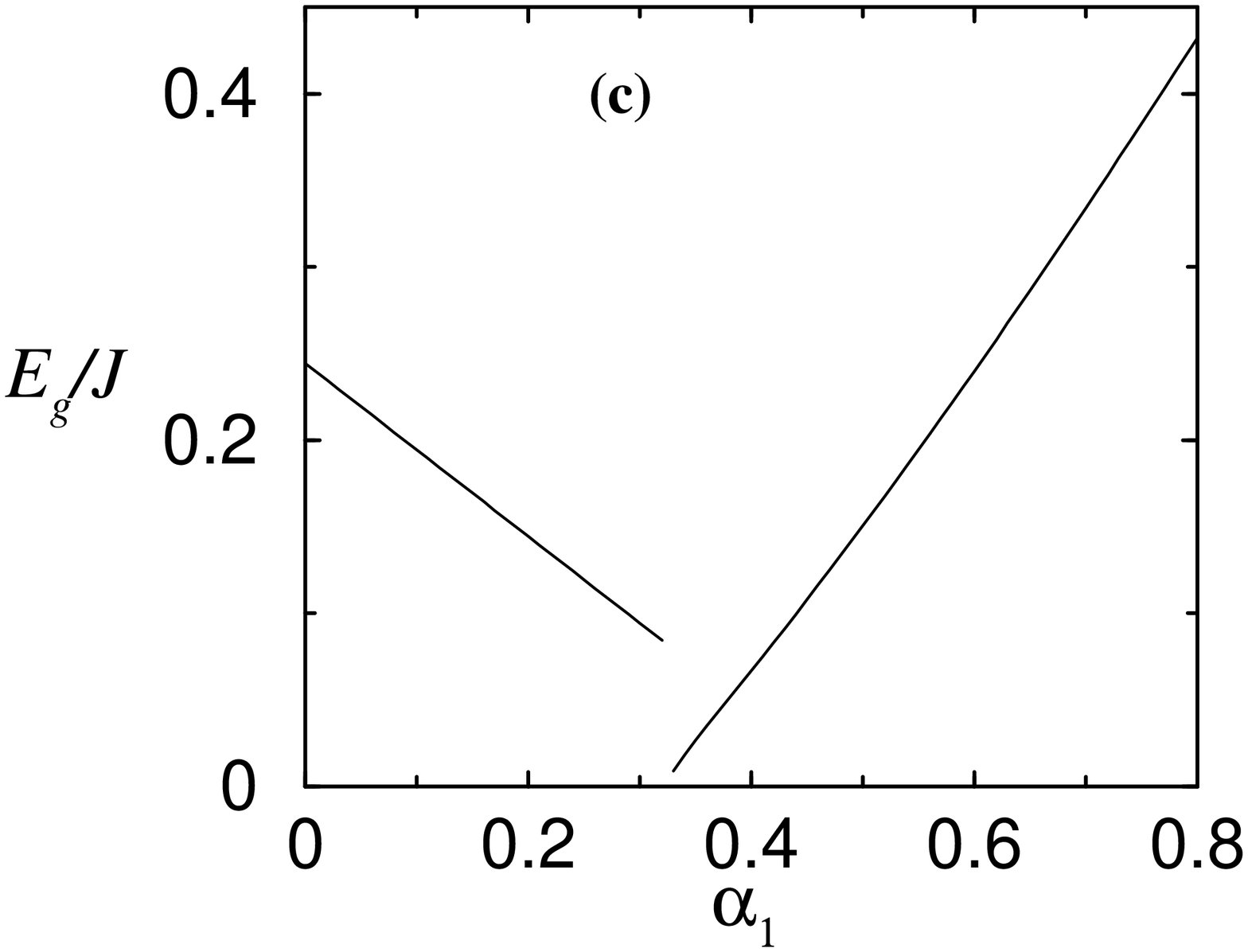}}
\caption{\label{fig:engap4} (a) The energy gap calculated for $\alpha_1= 1$ as a function of $\alpha_2$ is displayed. 
		The solid line is the data calculated using BMFT and the dotted one is the DMRG data from Ref.~\cite{Jd3}.
(b) The energy gap plotted for $\alpha_1= 0.5$ as a function of $\alpha_2$. The solid line is the data calculated using BMFT and the dotted one is the DMRG data from Ref.~\cite{Jd5}. (c) The energy gap is plotted as a function of $\alpha_1$ for $\alpha_2 = 0.2$.}
\end{center}
\end{figure}

Another quantity of significant importance is the energy gap, which characterizes all three states.
As a consequence all spin correlations span regions of size of the order of the reciprocal of the gap.
We calculated the gap as a function of $\alpha_2$ for $\alpha_1=1$ and reported it in Fig. \ref{fig:engap4}.
Our result is compared to the exact DMRG ones\cite{Jd3,Jd5} in the same figure. One can note that the gap we calculate is not in good quantitative agreement, but as far as trends are concerned good qualitative agreement is found. Our gap behaves linearly in all the mean-field calculations whereas in the DMRG it is nearly horizontal for small and large values of $\alpha_2$. The most important feature that both BMFT and DMRG results share is that the gap shows a minimum at the critical value of $\alpha_2$. In the DMRG results\cite{Jd3}, the transition was interpreted to be second order because the gap vanishes (within uncertainty) at the transition. In BMFT the transition is not gapless but the gap becomes very small. For $ \alpha_2 = 0.2$ the minimum gap value from DMRG\cite{Jd3} is $0.004J\pm0.004J$, and in BMFT it is $0.009J$. One should stress however that it is the crossing of the free (or ground-state) energies
that determines the transitions in BMFT not the vanishing of the energy gap.

\begin{figure}
\begin{center}
\subfigure{
\includegraphics[scale=0.22]{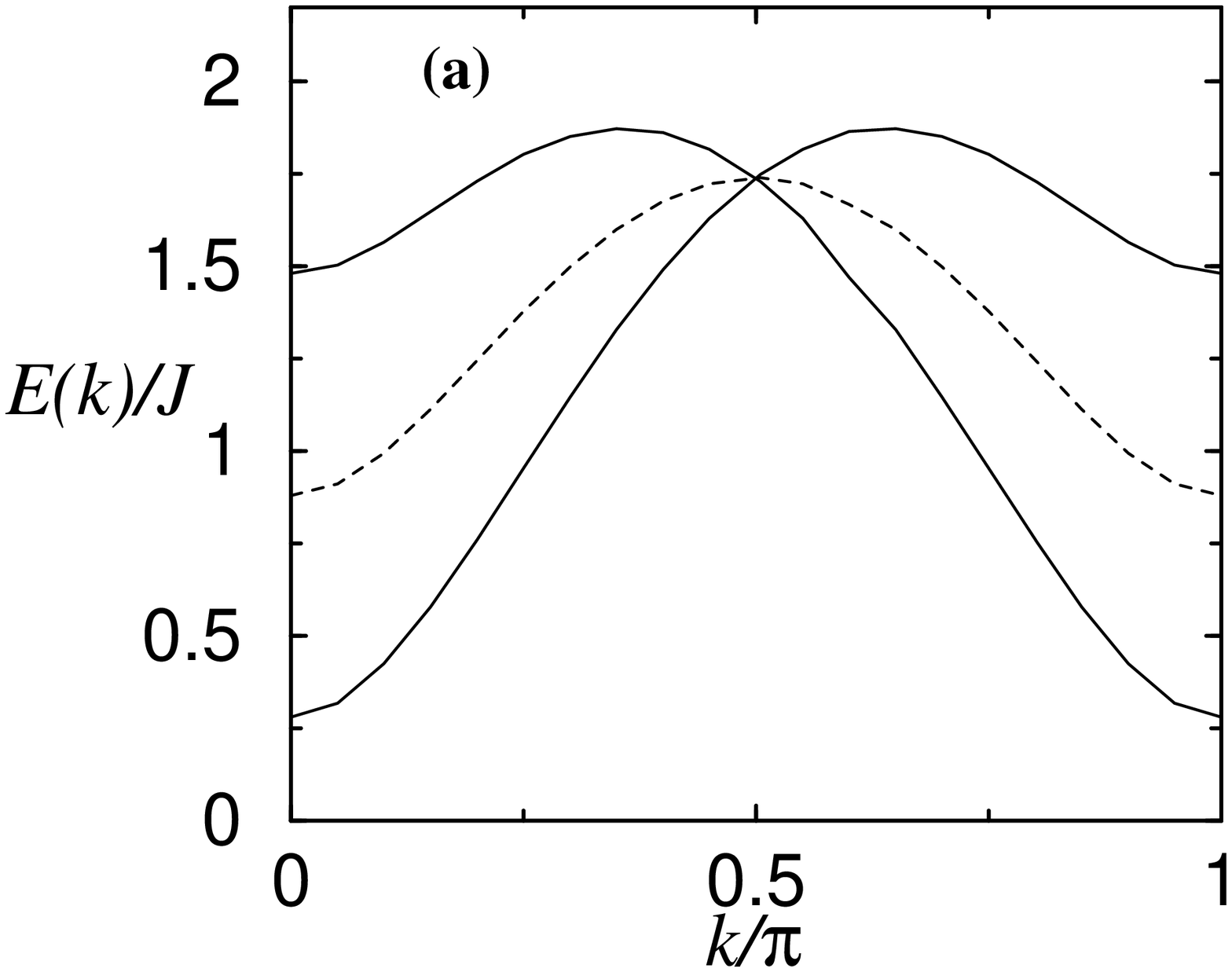}}
\subfigure{
\includegraphics[scale=0.22]{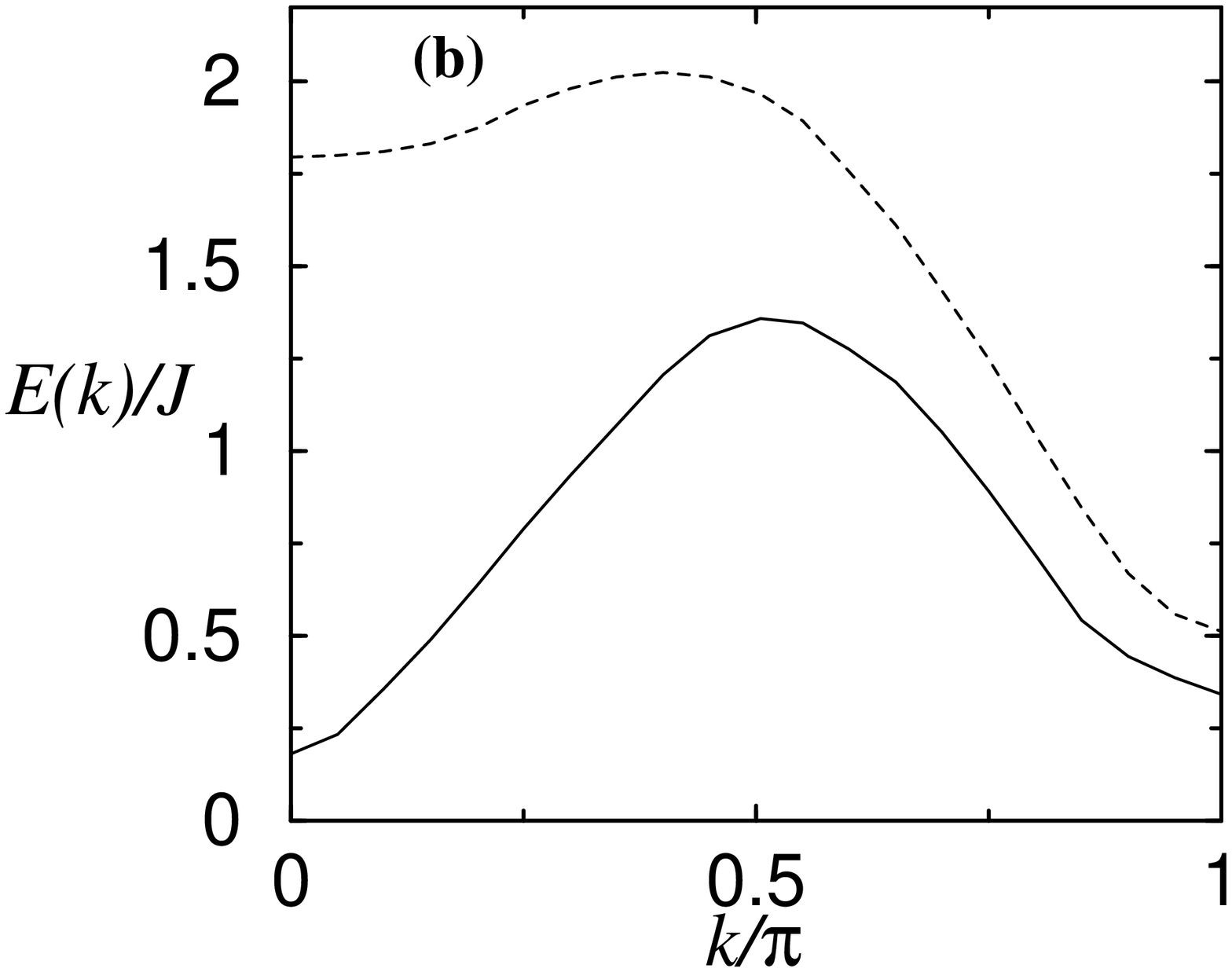}}
\caption{\label{fig:Nspectrum} (a) The energy spectra calculated within BMFT (dotted lines) are compared to those from the dimer expansion method of Ref.~\cite{Jd6}. (a) $\alpha_1 = 1$ and $\alpha_2=0$. (b) $\alpha_1 = 1$ and $\alpha_2=0.6$.} 
\end{center}
\end{figure}

The energy spectra for $\alpha_1= 1$ and $\alpha_2=0$, and for 
$\alpha_2=0.6$ both in the N-type state are shown in Fig.~\ref{fig:Nspectrum}. For $\alpha_1 = 1$ and zero diagonal coupling, the exact spectrum calculated numerically starts off asymmetric with a minimum at $k=\pi$.~\cite{Jd6,Barnes} As frustration increases the local minimum at 0 decreases rapidly but the absolute minimum at $\pi$ decreases only slowly. So the energy gap decreases slowly while the minimum is at $\pi$, but starts to decrease rapidly once the absolute minimum shifts to $k=0$. 

\begin{figure}
		\includegraphics[height=4.0cm]{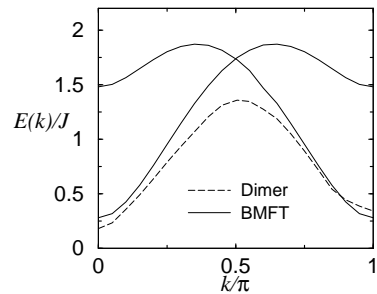}
		\centering
		\caption{Comparison of spectra for $\alpha_2$=0.6. The dashed curve is the result of the dimer expansion in Ref.\cite{Jd6}.}
	\label{fig:E(k)}
\end{figure}

Within the BMFT, the shape of the spectrum for $ \alpha_1 = 1$ and $ \alpha_2=0$ is different from the exact one. But as discussed in Ref.\cite{Azz5}, the important feature shared by both results is the presence of an energy gap. Also, BMFT rightfully describes the physics of the two-leg ladder in this limit, namely that the ground state consists of the formation of renormalized spin singlets on the rungs. The difference in the curves for the energy gap (Fig.~\ref{fig:engap4}) is due to the complex behavior of the spectra shape as the couplings are varied. Near the boundary line between R and N-type states, for $\alpha_1 = 1$ and $\alpha_2=0.6$ the BMFT spectra yield a low-lying excitation spectrum that overall behaves like the dimer-expansion data of Ref.~\cite{Jd6}; see Fig.~\ref{fig:E(k)}.

\subsection{Nonzero-temperature phase diagram}

\begin{figure}[b]
		\includegraphics[height=4.0cm]{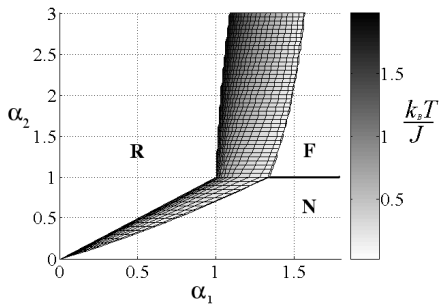}
		\centering
		\caption{Surface plot of the phase diagram showing the temperature dependence.}
	\label{fig:PhaseTemp}
\end{figure}

Unlike exact diagonalization methods, the present analytical approach can be readily used to analyze the effects of temperature on the system. We repeated the same approach as in Sec. \ref{sec:zerotemp} by comparing the free energies of the three phases, this time, at different temperatures for various sets of coupling values. We deduced the temperature dependence of the phase boundaries in the phase diagram. The result of such a calculation is reported in Fig.~\ref{fig:PhaseTemp}. We found that as temperature increases the R-type state decreases in size. The sizes of the N-type and F-type phases increase with temperature.

In the R-type state at zero temperature, the spins on the rungs arrange themselves ferromagnetically; notice that on any rung the pair of spins fluctuate together between the up and down spin orientations while the two pairs of spins on the adjacent rungs fluctuate in the opposite direction. So, the system is neither ordered ferromagnetically
nor antiferromagnetically. As mentioned earlier, the parameter $P$ is a measure of the AF correlations along the rung direction. Because of the ferromagnetic orientation, their AF correlations are zero, leading to $P=0$. This parameter becomes non-zero as temperature increases as seen in Fig.~\ref{fig:ParametersTemp} because thermal fluctuations allow the rungs to adopt sometimes the AF arrangement. Note that eventually, the AF correlations diminish in the very high temperature limit, a fact that is indicated by $P$ decreasing as $1/T$ after reaching a maximum. In the Ising limit, the phase boundary between the R-type and N-type phases is $\alpha_1= 2\alpha_2$, and the boundary between the R-type and and F-type phases is $\alpha_1=2$. Quantum fluctuations cause these phase boundaries to move toward the R-type state. Including thermal fluctuations seems to have the same effect. It is found that as temperature increases the phase boundaries move toward $\alpha_1 = \alpha_2$ and $\alpha_1=1$, respectively. As temperature rises $P$ approaches but never becomes larger than $P'$ and $Q$ for $ \alpha_1= \alpha_2=1$, Fig. \ref{fig:ParametersTemp}.

\subsection{High-temperature regime}

\begin{figure}
\begin{center}
\subfigure{
\includegraphics[scale=0.23]{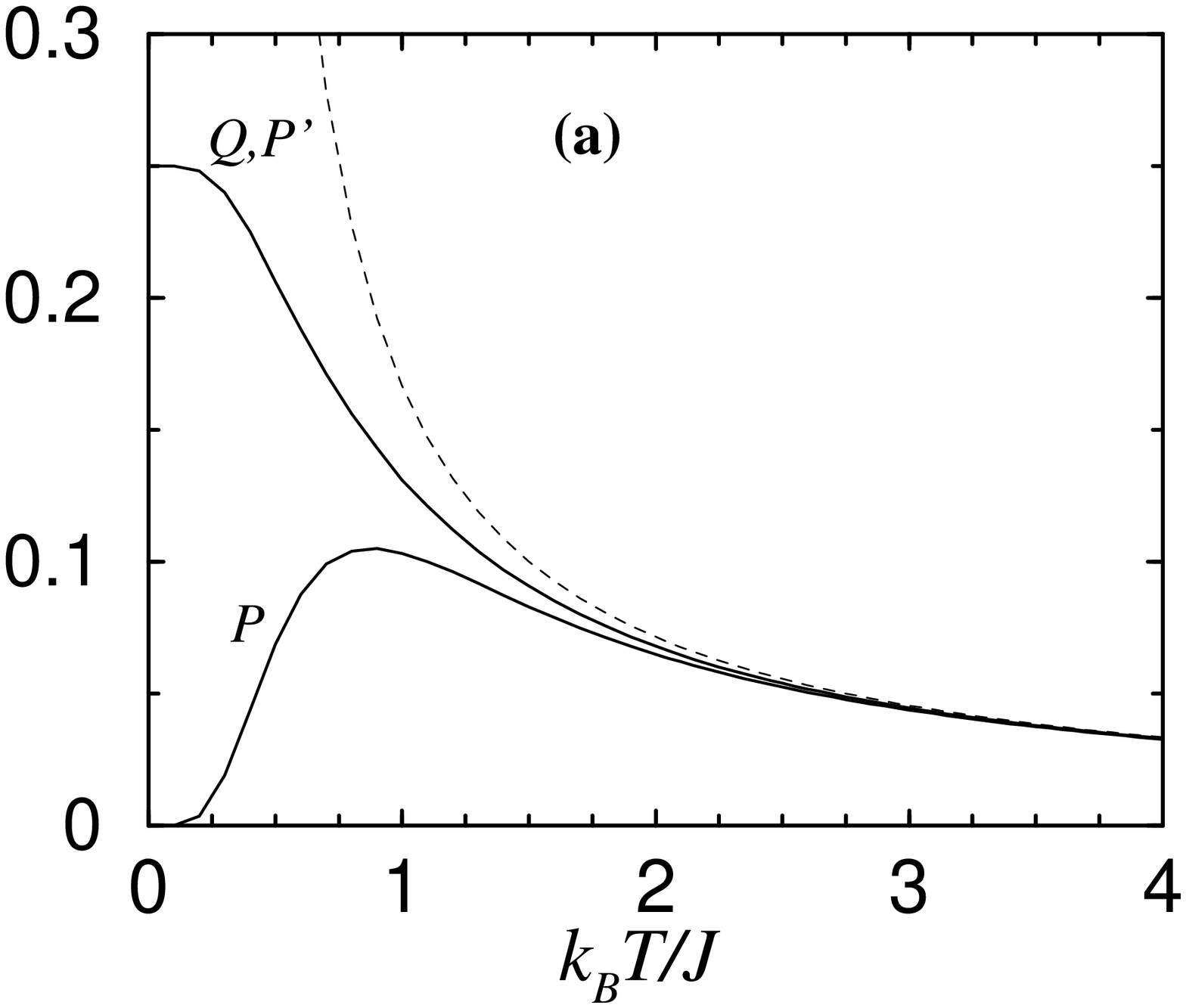}}
\subfigure{
\includegraphics[scale=0.23]{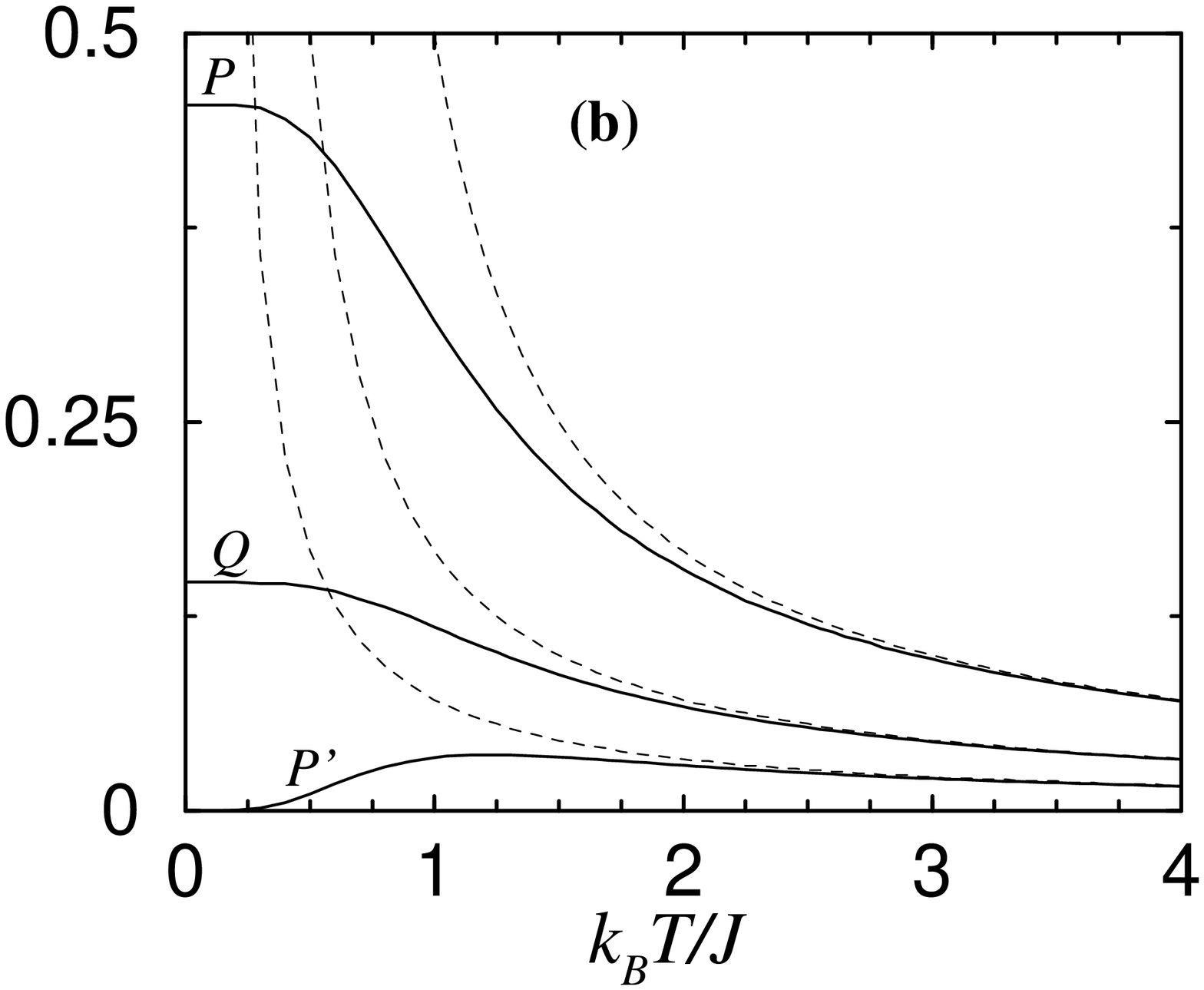}}
\caption{\label{fig:ParametersTemp}The mean-field parameters are plotted versus temperature (solid lines). The dotted lines are from the high-temperature limit equations~(\ref{eq:HT}). (a) $ \alpha_2= \alpha_1=1$ in the R-type state. (b) $ \alpha_2=0.5$ and $ \alpha_1=2$ in the N-type state.}
\end{center}
\end{figure}
In the limit $k_{B}T \gg J$, $J_{\perp}$ and $J_{\times}$ the mean-field equations can be solved analytically. The Fermi-Dirac factors ($\tanh$ functions) can be expanded to first order in $\beta E_{p}$ in the mean-field equations.\cite{Azz5} The approximation is subbed into Eqs.~(\ref{eq:sc}) with $\sum_{k} \to \int \frac{dk}{2\pi}$, and the following set of equations are obtained
\begin{eqnarray}
\label{eq:HT}
Q & \approx & \frac{J}{8k_{B}T(1-\frac{J}{4k_{B}T})}, \ \ \ k_{B}T \gg J,\nonumber \\
P & \approx & \frac{J_{\perp}}{8k_{B}T(1-\frac{J_{\perp}}{4k_{B}T})}, \ \ \ k_{B}T \gg J_{\perp}, \nonumber \\
P' & \approx & \frac{J_{\times}}{8k_{B}T(1-\frac{J_{\times}}{4k_{B}T})}, \ \ \ k_{B}T \gg J_{\times}.
\label{highT}
\end{eqnarray}
We find that these equations are independent of the state in which they are calculated; i.e., whether we use the set of equations for the N-type, R-type, or F-type state we always get the same result (\ref{highT}) in the high-$T$ regime. In this regime, the parameters decrease following a Curie-Weiss $T^{-1}$ law but never vanish, excluding in this way the occurrence of any finite-temperature phase transition from a state with finite spin bond order to a high-$T$ state with zero spin bond order. Note that Eqs. (\ref{highT}) fit very well the numerically calculated parameters as seen in Fig.~\ref{fig:ParametersTemp}. It is interesting to note  that all the parameters have the same form and show the same dependence on the ratio of the coupling constant, in the direction in which the parameter is calculated, and temperature. Note that the smallest parameter corresponds to the direction in which the spins are ferromagnetically arranged, e.g. $P$ is the smallest parameter in the R-type state. At high temperature, it is easy to see from~(\ref{eq:HT}) that the parameter corresponding to the smallest coupling value will be smallest. So, in the high-temperature limit the largest coupling value determines the state of the system. This is why the boundaries in the phase diagram shift to $\alpha_1=\alpha_2$ and $\alpha_1=1$.

\section{Criticality at nonzero temperature}
\label{sec:Tcriticality}

\begin{figure}
\begin{center}
\subfigure{
\includegraphics[scale=0.23]{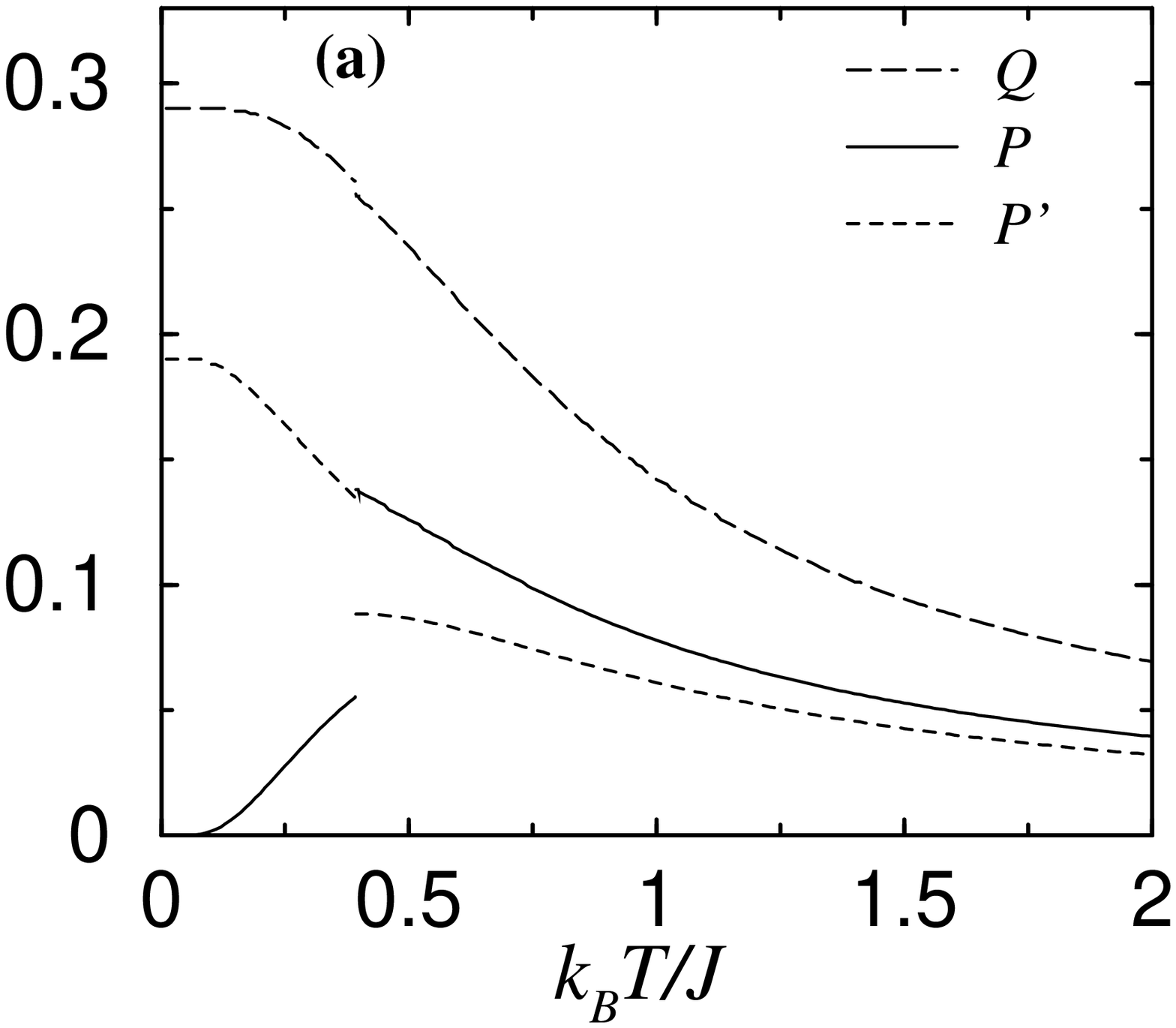}}
\subfigure{
\includegraphics[scale=0.23]{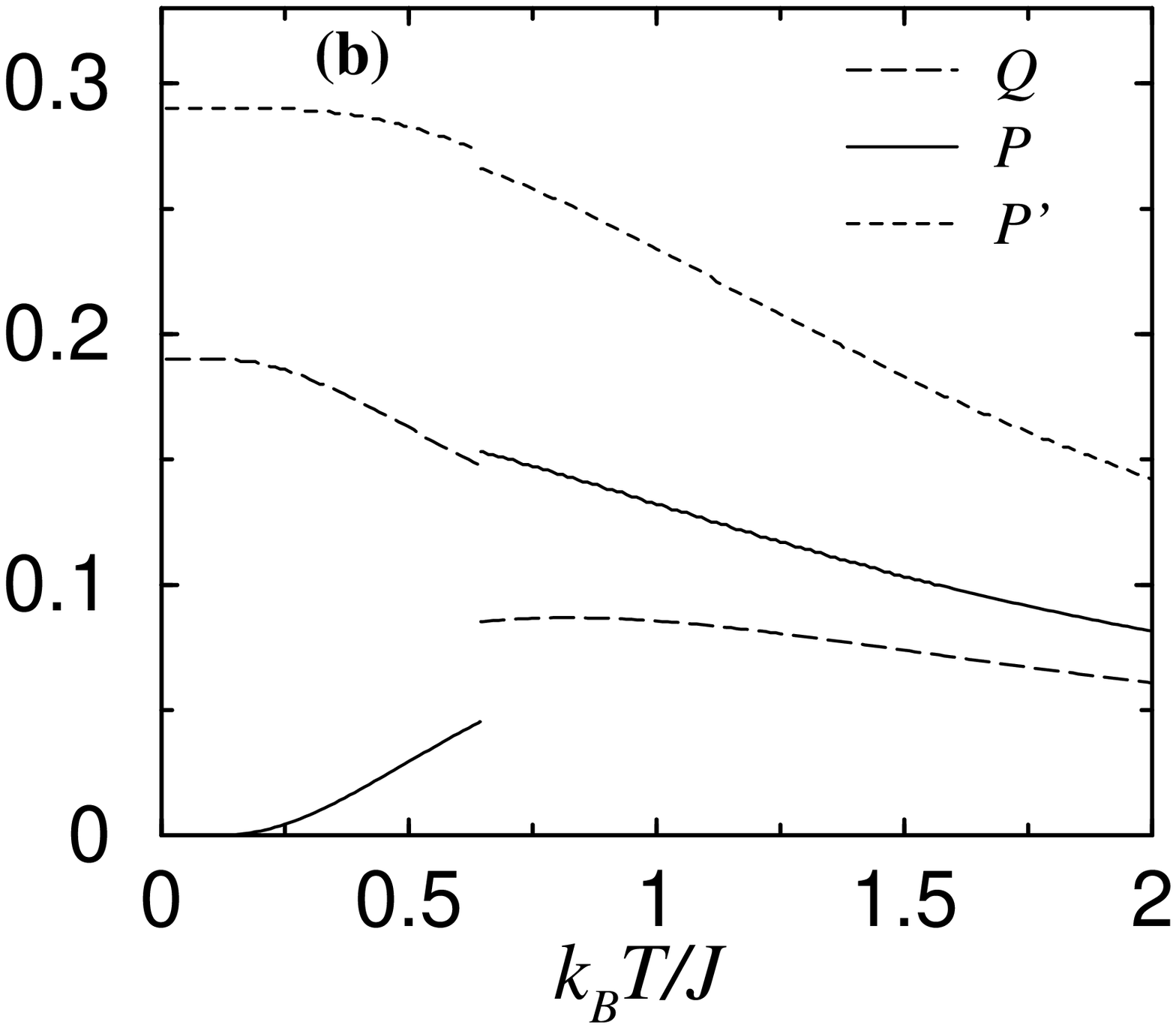}}
\caption{\label{fig:ParametersTemp2}The parameters are plotted versus temperature for two coupling sets.
(a) $ \alpha_2=0.5$ and $ \alpha_1=0.6$. The phase transition happens at $k_{B}T/J=0.39$ from the R-type to the N-type state.(b) $ \alpha_2=2$ and $ \alpha_1=1.25$. The phase transition happens at $k_{B}T/J=0.64$ from the R-type to the F-type state.}
\end{center}
\end{figure}

\begin{figure}
\begin{center}
\subfigure{
\includegraphics[scale=0.22]{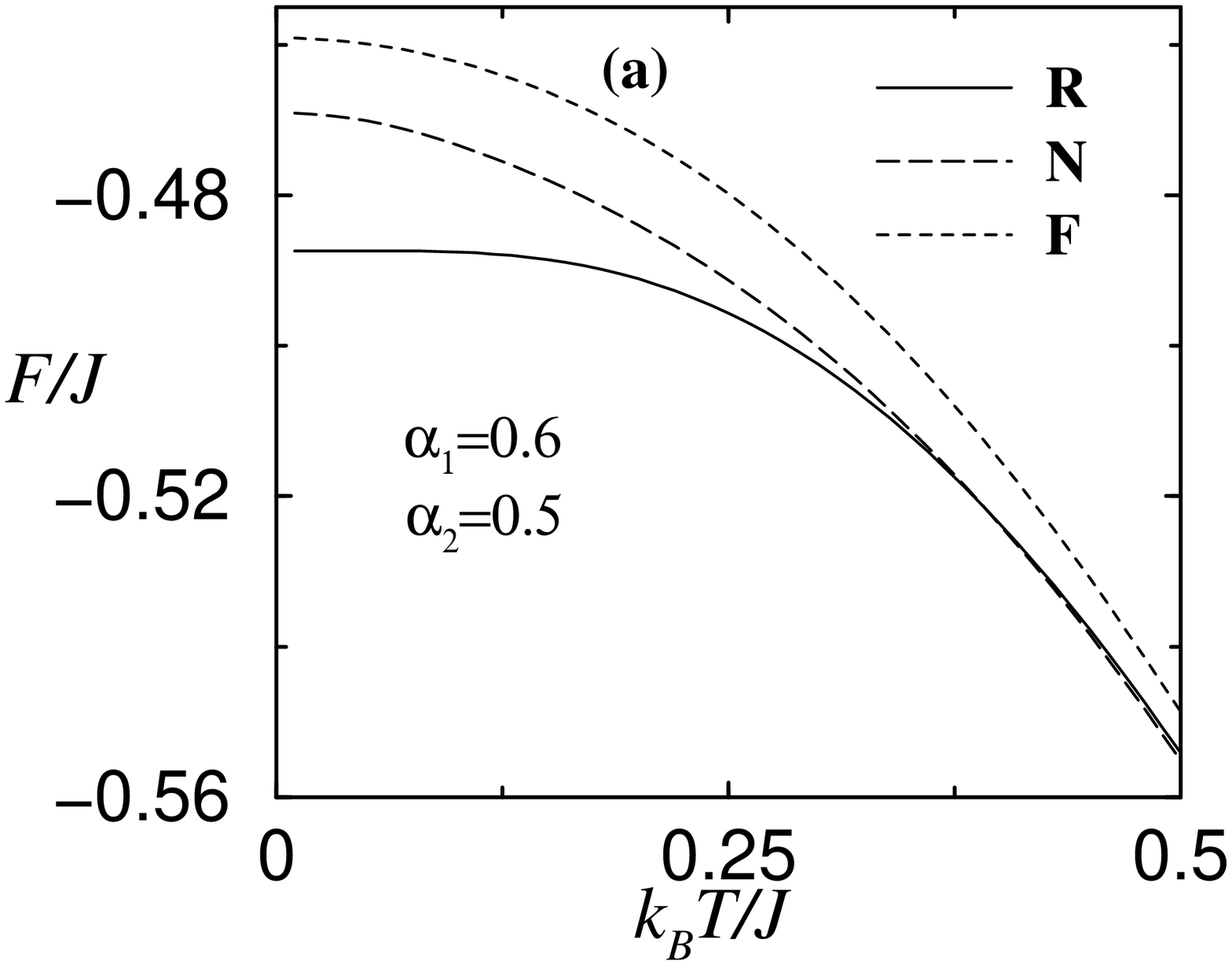}}
\subfigure{
\includegraphics[scale=0.22]{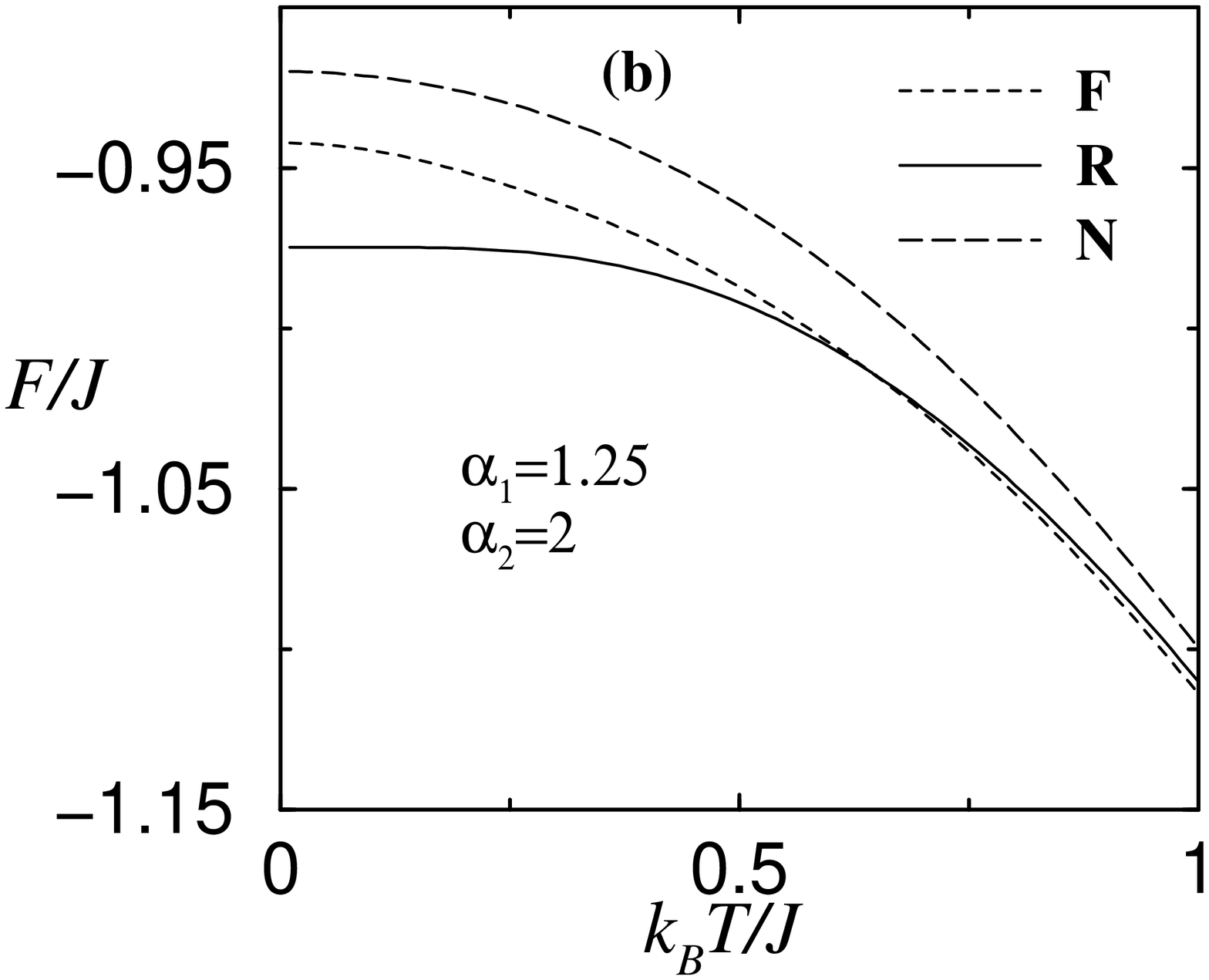}}
\caption{\label{fig:freeEnergy2}The free energies are plotted as functions of temperature. The phase transitions happen where the free energies cross.
(a) $ \alpha_2=0.5$ and $ \alpha_1=0.6$. The phase transition happens at $k_{B}T/J=0.39$ from the R-type to the N-type state.(b) $ \alpha_2=2$ and $ \alpha_1=1.25$. The phase transition happens at $k_{B}T/J=0.64$ from the R-type to the F-type state.}
\end{center}
\end{figure}

\begin{figure}
\begin{center}
\subfigure{
\includegraphics[scale=0.22]{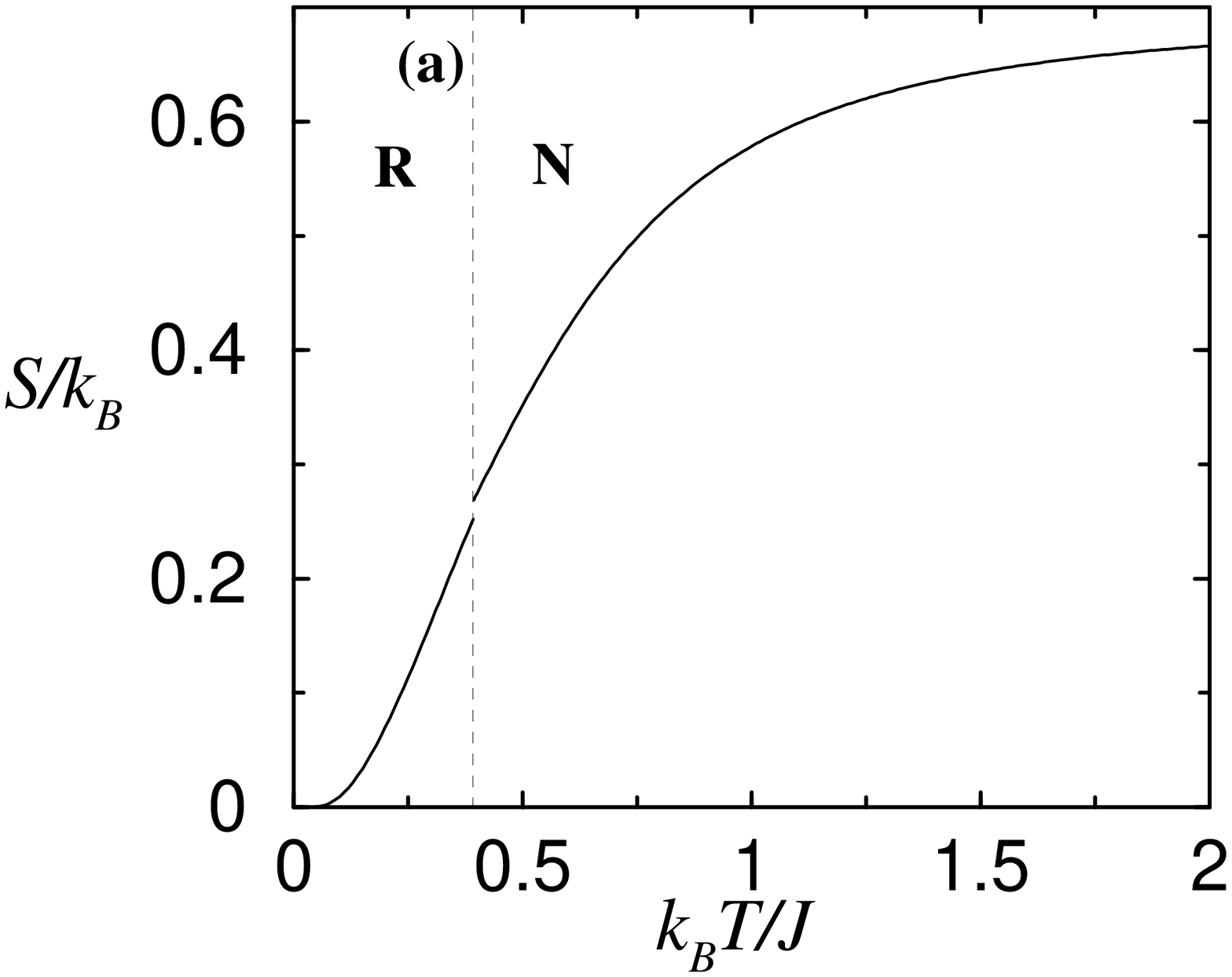}}
\subfigure{
\includegraphics[scale=0.22]{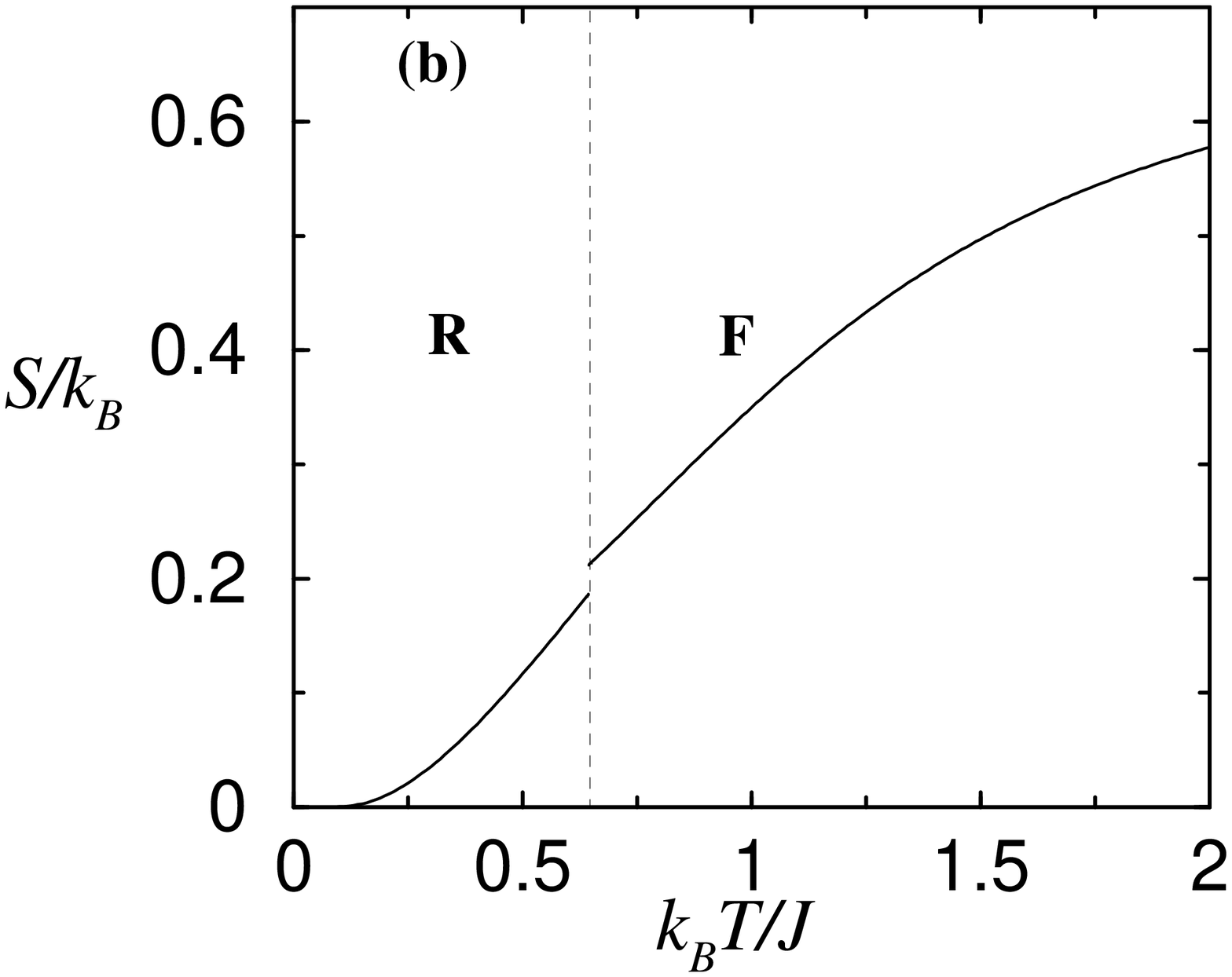}}
\caption{\label{fig:EntropyTemp2}The entropy is plotted versus temperature. The entropy is discontinuous at the transition points. The dashed lines indicate the transition between the different phases.
(a) $ \alpha_2=0.5$ and $ \alpha_1=0.6$. The phase transition happens at $k_{B}T/J=0.39$ from the R-type to the N-type state. (b) $ \alpha_2=2$ and $ \alpha_1=1.25$. The phase transition happens at $k_{B}T/J=0.64$ from the R-type to the F-type state.}
\end{center}
\end{figure}

\begin{figure}
\begin{center}
\subfigure{
\includegraphics[scale=0.22]{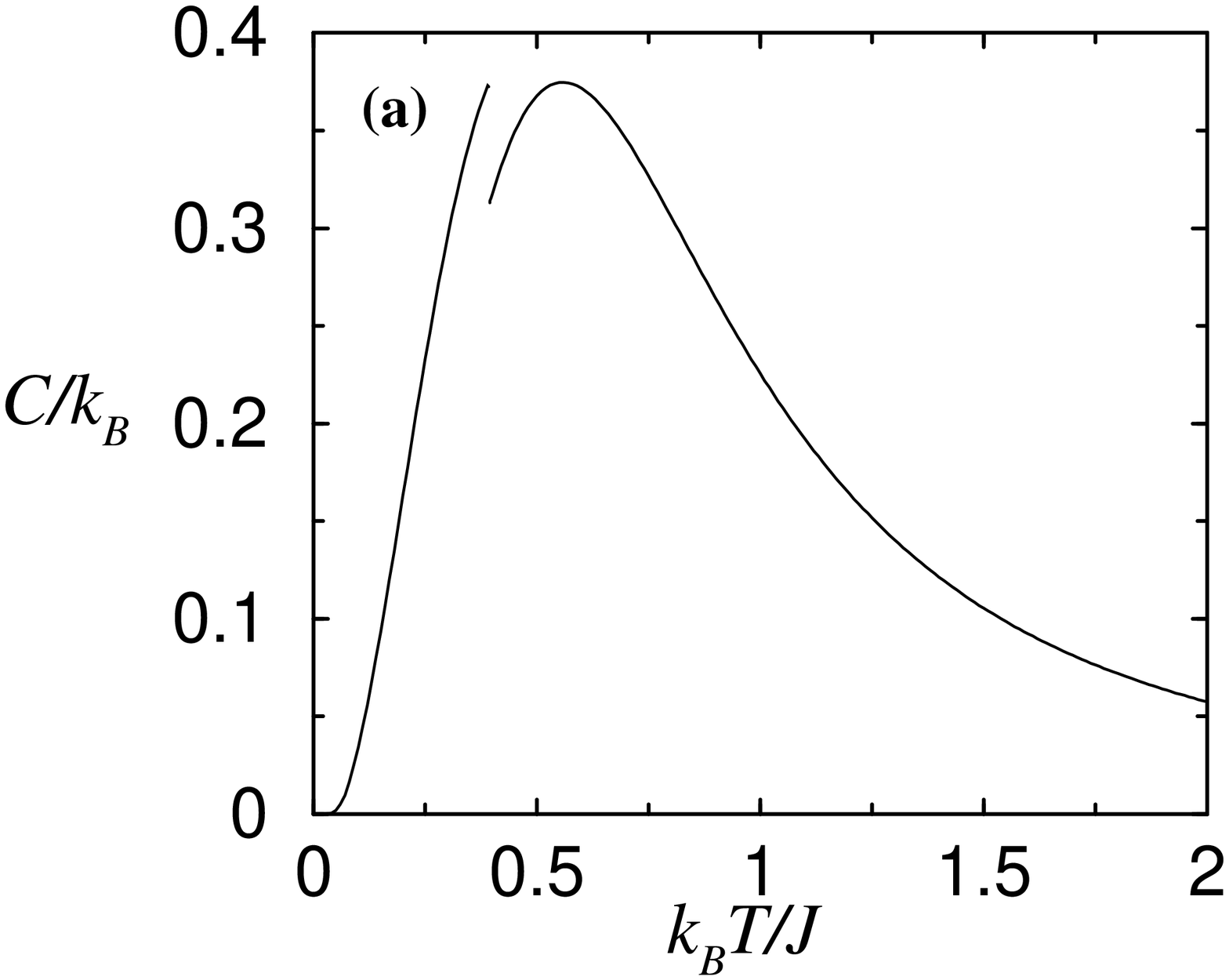}}
\subfigure{
\includegraphics[scale=0.22]{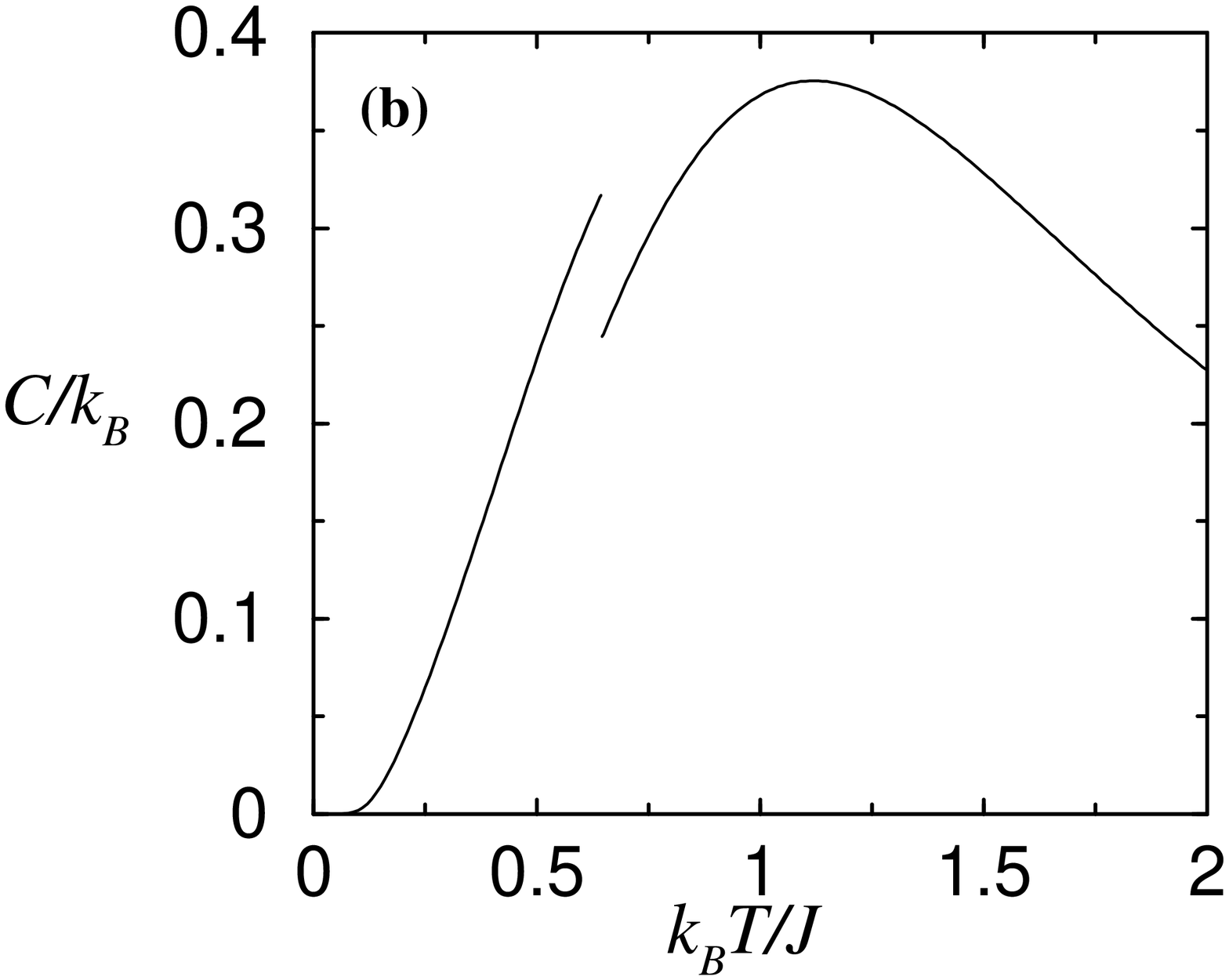}}
\caption{\label{fig:HeatCapacity}The specific heat is calculated by $C=T\pderiv{S}{T}$. The infinite peaks at the transitions, which are a consequence of the discontinuity in $S(T)$, are not displayed for clarity. (a) $ \alpha_2=0.5$ and $ \alpha_1=0.6$. The phase transition happens at $k_{B}T/J=0.39$ from the R-type to the N-type state. (b) $ \alpha_2=2$ and $ \alpha_1=1.25$. The phase transition happens at $k_{B}T/J=0.64$ from the R-type to the F-type state.}
\end{center}
\end{figure}

For sets of coupling values ($\alpha_{1}$, $\alpha_2$) within the shaded region of Fig.~\ref{fig:PhaseTemp}, the thermal fluctuations can cause a first-order phase transition from the R-type state to the F-type state or N-type state. There are no thermally-induced transitions between the F-type and N-type states because the boundary between the N and F-type states is not temperature dependent due to the symmetry of the Hamiltonian with respect to exchanging $J$ and $J_\times$ terms. 
Note that thermally-induced transitions are not from a 
disordered phase to an ordered one, or vice versa.
The mean-field bond parameters are displayed in 
Fig.~\ref{fig:ParametersTemp2} as functions of temperature for 
two sets of couplings $\alpha_1$ and $\alpha_2$ such that Fig.~\ref{fig:ParametersTemp2}a shows a transition from the R-type to the N-type state and Fig.~\ref{fig:ParametersTemp2}b shows a transition from the R-type to the F-type state.
The corresponding free 
energies are shown in Fig.~\ref{fig:freeEnergy2}, entropies in 
Fig.~\ref{fig:EntropyTemp2}, and specific heats in 
Fig.~\ref{fig:HeatCapacity} for the same sets of couplings.

The entropy is calculated using
\begin{eqnarray}
\label{eq:Entropy}
S&=&-\frac{k_B}{2N} \sum_{k} \sum_{p=1}^{4} 
\big\{ n_{F}[E_{p}(k)]\ln \{n_{F}[E_{p}(k)]\}\nonumber \\ 
&&+ \{1-n_{F}[E_{p}(k)]\} \ln \{1-n_{F}[E_{p}(k)]\} \big\}
\end{eqnarray}
which is derived from $S=-\pderiv{F}{T}$. In Eq.~(\ref{eq:Entropy}), $E_{p}(k)$ refers to the energy spectra of the state where $S$ is calculated. The specific 
heat is calculated using $C=T\pderiv{S}{T}$. 
The entropy shows a discontinuity at $T_C$, implying 
that the transition is first-order in character.
At very high temperature entropy saturates as expected to a value of $k_{B}\ln2$.
In Fig. \ref{fig:EntropyTemp2}(a) for $ \alpha_2=0.5$ and $ \alpha_1=0.6$, the phase transition happens at $k_{B}T/J=0.39$ from the R-type to the N-type state when $T$ increases. In Fig. \ref{fig:EntropyTemp2}(b) with $\alpha_2=2$ and $ \alpha_1=1.25$, the phase transition happens at $k_{B}T/J=0.64$ from the R-type to the F-type state.
In these figures, the dashed lines simply indicate the transitions between the different phases corresponding to the set of couplings used.
We found that in the limit $\alpha_1$ and $\alpha_2\to 0$ the 
jump in entropy goes to zero.  For small coupling values 
this jump could be smaller than experimental precision 
(if a real material existed) so that it would become difficult to 
assert that the transition is a first-order one. 
For example, for $ \alpha_2 = 0.2$ 
and $\alpha_1 = 0.24$ the jump in entropy is about $0.002k_{B}$ only. Note that because all three states are gapped, both entropy and specific heat show an activated behaviour in the viscinity of zero temperature.

All the phases of the system 
are disordered; i.e., none of them is characterized by 
long-range magnetic order of any kind. 
The proposed thermally-induced criticality can be seen as a remnance of the zero-$T$ 
(quantum) criticality because of the temperature dependence of the phase 
boundaries.
As mentionned earlier in the introduction, the mean-field parameters $Q$, $P$, and $P'$ do not represent any kind of long-range order. The zero-temperature phase transitions we analyzed in Sec. \ref{sec:zerotemp} all occur between disordered phases that differ only by the way the spins arrange themselves on very short distances (refer to Fig.~\ref{fig:Jd1}). Therefore, we believe that the finite-$T$ transitions we find here for the present practically one-dimensional system are not an artifact of the mean-field character of BMFT. These phase transitions are due to frustration; i.e., they disappear once frustration is brought to zero. They are also a consequence of the fact that the zero-$T$ quantum phase transition boundary depends on temperature.
Because zero temperature cannot be reached in practice, a quantum phase transition cannot in fact be observed directly. For the present system, the signature for such a transition would be the observation of the finite-$T$ transitions (if an experimental system existed).
\section{Discussion}
\label{sec:Discussion}

While the quantum criticality we find in this system has been already found elsewhere using exact numerical methods, the criticality at finite temperature we report on here remains to be confirmed by other theoretical methods. Experimentally, if ever a frustrated two-leg ladder where the strength of frustration is as important as the coupling along the rungs and chains existed, then thermodynamical measurements would either confirm or refute our claims. Perhaps the application of a high pressure on a two-leg ladder material could increase the diagonal interaction and allow the search for this thermally-induced criticality. In the absence of diagonal interaction (frustration), the AF Heisenberg two-leg ladder shows neither quantum nor classical criticality. Interestingly, it is the frustration that is responsible for both criticalities. A strong argument in favor of the existence of the finite-$T$ criticality is the existence of the zero-$T$ one itself because in both cases each phase boundary separates two of the same three disordered phases. Denying the finite-$T$ criticality would amount to denying the zero-$T$ one according to the present theory.

\section{Conclusion}
\label{sec:Conclusion}

In this work, we studied quantum as well as classical criticality in the 
two leg-ladder with exchange interactions along the chains, rungs, and diagonals 
using the Jordan-Wigner transformation and the bond-mean-field theory. 
The zero-temperature phase diagram of this system is calculated. We 
found it to exhibit three quantum phases, characterized all by an 
energy gap and absence of magnetic order. These states are labeled 
N\'eel-type (N-type), Rung-type (R-type or Haldane-type) and 
Ferromagnetic-type chain (F-type) states. This result agrees well 
with existing numerical data. The transitions from one phase to any 
of the two others are all first-order. 
Because they occur at zero 
temperature, they enter under the category of quantum phase transitions. 
When temperature increases for some sets of coupling values, the system 
undergoes a phase transition from the R-type state to the 
N or F-type state at a finite temperature. The finite temperature 
phase diagram is calculated as well. In it, the size of the R-type 
state becomes smaller while the F-type state and the N-type state 
increase in size with increasing temperature. The good agreement 
between our results and existing exact results for the zero-$T$ 
phase diagram suggests that the present mean-field treatment is 
acceptable. The various phase transitions found in 
this work occur between magnetically disordered states. It is 
solely the frustration, the spin bond parameters and the nature 
of short-range magnetic correlations that determine the nature 
of all three phases that characterize the frustrated antiferromagnetic 
Heisenberg two-leg ladder.
In the present mean-field-type approach, all phase transitions 
are first-order in character. While at zero temperature existing exact 
numerical data
seem to indicate that this is the case, at finite-temperature 
the degree of the transitions
remains to be confirmed.

\begin{acknowledgments}
We wish to acknowledge the financial support of the Natural 
Science and Engineering Research Council of Canada
(NSERC).
\end{acknowledgments}


\begin{thebibliography}{5}


\bibitem{sachdev1999}S. Sachdev, in \textit{Quantum Phase Transitions}, Cambridge University Press, New York, 1999.

\bibitem{Dagotto}For a review, see E. Dagotto and T.M. Rice, Science \textbf{271}, 618 (1996).

\bibitem{Azuma}M. Azuma, Z. Hiroi, M. Takano, K. Ishida, and Y. Kitaoka, Phys. Rev. Lett. \textbf{73}, 3463 (1994).

\bibitem{Hayward}C.A. Hayward, D. Poilblanc, and L.P. L\'evy, Phys. Rev. B \textbf{54}, R12649
(1996).

\bibitem{Chaboussant}G. Chaboussant, P.A. Crowell, L.P. L\'evy, O. Piovesana, A. Madouri, and D. Mailly, Phys. Rev. B \textbf{55}, 3046 (1997).

\bibitem{Imai}T. Imai, K. R. Thurber, K.M. Shen, A.W. Hunt,  and F.C. Chou, Phys. Rev. Lett. \textbf{81}, 220 (1998).

\bibitem{Jd11}M. Matsuda, K. Katsumata, R.S. Eccleston, S. Brehmer, and H.-J. Mikeska, J. Applied Phys. \textbf{87}, 6271 (2000).

\bibitem{Jd6}Z. Weihong, V. Kotov, and J. Oitmaa, Phys. Rev. B \textbf{57}, 11439 (1998).

\bibitem{Jd4}T. Sakai and N. Okazaki, Journal of Applied Physics \textbf{87}, 5893 (2000).

\bibitem{Jd8}H.-H. Hung, C.-D. Gong, Y.-C. Chen, and M.-F. Yang, cond-mat/0605719, (2006).

\bibitem{white1996} S.R. White, Phys. Rev. B  \textbf{53}, 52 (1996).

\bibitem{Jd10} T. Hakobyan, J.H. Hetherington, and M. Roger, Phys. Rev. B \textbf{63}, 144433 (2001).

\bibitem{Jd3}X.Q. Wang Mod. Phys. Lett. B \textbf{14}, 327 (2000).
\bibitem{Jd5} N. Zhu, X. Wang, and C. Chen, Phys. Rev. B \textbf{63}, 012401 (2000).

\bibitem{Jd7} Y. Xian, Phys. Rev. B \textbf{52}, 12485 (1995).

\bibitem{Jd13}D. Allen, F.H.L. Essler, and A.A. Nersesyan, Phys. Rev. B \textbf{61}, 8871 (2000).

\bibitem{hakobyan2007} T. Hakobyan, cond-mat/0702148 (2007).

\bibitem{starykh2004} O.A. Starykh and L. Balents, Phys. Rev. Lett. {\bf 93}, 127202 (2004).

\bibitem{Jd12}C-M Nedelcu, A. K. Kolezhuk, and H-J Mikeska, J. Phys.:Condens. Matter \textbf{12}, 959 (2000).

\bibitem{Jd9}D.C. Cabra, A. Dobry, and G.L. Rossini, Phys. Rev. B \textbf{63}, 144408 (2001).

\bibitem{Azz3}M. Azzouz, L. Chen, and S. Moukouri, Phys. Rev. B \textbf{50}, 6233 (1994).

\bibitem{Azz2}B. Bock and M. Azzouz, Phys. Rev. B \textbf{64}, 054410 (2001).

\bibitem{Azz1}M. Azzouz, Phys. Rev. B \textbf{48}, 6136 (1993).

\bibitem{Azz5}M. Azzouz, Phys. Rev. B \textbf{74}, 174422 (2006).

\bibitem{kim2000} E.H. Kim, G. F\`ath, J. S\'olyom, and D.J. Scalapino, Phys. Rev. B {\bf 62}, 14965 (2000).

\bibitem{Mermin}N.D. Mermin and H. Wagner, Phys. Rev. Lett. \textbf{17}, 1133 (1966).

\bibitem{Affleck}I. Affleck and J.B. Marston, Phys. Rev. B \textbf{37}, 3774 (1988).

\bibitem{Cloiseaux}J. des Cloiseaux and J.J. Pearson, Phys. Rev. \textbf{128}, 2131 (1962).

\bibitem{XiDai}Xi Dai and Zhao-bin Su, Phy. Rev. B \textbf{57}, 964 (1998).

\bibitem{Barnes}T. Barnes, E. Dagotto, J. Riera, and E. S. Swanson, Phys. Rev. B \textbf{47}, 3196 (1993).












\end{thebibliography}
\end{document}